\documentclass[aps,pre,twocolumn,superscriptaddress]{revtex4-1}
\usepackage[english]{babel}
\usepackage{graphicx}

\usepackage{amsmath}
\usepackage{amsfonts}
\usepackage[usenames,dvipsnames]{color}

\usepackage{hyperref}
\usepackage{rotating}
\usepackage{booktabs} 
\usepackage{slashed}
\usepackage{transparent}
\usepackage{color}
\usepackage{soul}
\usepackage{tikz}
\usetikzlibrary{arrows}

\usetikzlibrary{patterns}
\usetikzlibrary{shapes,snakes}

\newcommand{\dd}{\; \mathrm{d}}
\newcommand{\va}[1]{{\boldsymbol{r}_{#1}}}

\DeclareMathAlphabet{\mathbb}{U}{bbold}{m}{n}

\newcommand{\bs}[1]{\boldsymbol{#1}}

\usepackage[normalem]{ulem}

\usepackage{hyperref}
\usepackage{nameref}

\begin{document}

\title{Continuum percolation expressed in terms of density distributions}

\author{Fabian Coupette}
\email{fabian.coupette@physik.uni-freiburg.de}
\author{Andreas H\"artel}
\author{Tanja Schilling}
\affiliation{Institute of Physics, University of Freiburg, Hermann-Herder-Stra{\ss}e 3, 79104 Freiburg, Germany}

\date{\today}

\begin{abstract}
We present a new approach to derive the connectivity properties of pairwise interacting n-body systems in thermal equilibrium. We formulate an integral equation that relates the pair connectedness to the distribution of nearest neighbors. For one-dimensional systems with nearest-neighbor interactions, the nearest-neighbor distribution is, in turn, related to the pair correlation function $g$ through a simple integral equation. As a consequence, for those systems, we arrive at an integral equation relating $g$ to the pair connectedness, which is readily solved even analytically if $g$ is specified analytically. We demonstrate the procedure for a variety of pair-potentials including fully penetrable spheres as well as impenetrable spheres, the only two systems for which analytical results for the pair connectedness exist. However, the approach is not limited to nearest-neighbor interactions in one dimension. Hence, we also outline the treatment of external fields and long-ranged interactions, and we illustrate how the formalism can applied to higher-dimensional systems using the three-dimensional ideal gas as an example.
\end{abstract}

\maketitle

\section{\label{sec:Introduction}Introduction}

Clustering of particles into connected aggregates is a process that occurs frequently in nature as well as 
in materials processing. The conditions under which a cluster becomes system spanning are of particular 
technological interest, as such a cluster might support mechanical stress (\textit{e.g.}~in the case of gels) or 
transport charges (\textit{e.g.}~in the case of conductive particles immersed in an insulating host 
matrix). The parameters associated to the emergence of a system-spanning cluster define the percolation threshold \cite{{bollobas2006percolation,stauffer2018introduction}}, the calculation of which for different systems has been subject of a vast number of studies. Accordingly, a rich methodology has evolved ranging from simulation \cite{{seaton1987aggregation,lee1988pair,rintoul1997precise,miller2003competition,consiglio2003continuum}} over liquid state theory \cite{{xu1988analytic,coniglio1977pair,desimone1986theory,kyrylyuk2008continuum,chiew1989connectivity,chatterjee2000continuum}} to renormalization group techniques \cite{cardy1996scaling}, stochastic tools \cite{{meester1996continuum,grimmett1999percolation}}, and conformal field theory \cite{smirnov2001critical,cardy1992critical}.  Nevertheless, exact (and non-trivial) results are only known for a couple of discrete systems, \textit{e.g.} random graphs \cite{bollobas2001random} or specific two-dimensional lattice systems \cite{{sykes1964exact,smirnov2001critical2}}.

 In disordered systems such as complex liquids, the interplay between liquid structure and connectivity is non-trivial.
 As a consequence, theories that are general enough to be applicable to a variety of different systems but still allow for immediate computation and sensible estimates of the key quantities, are rare. Thus, the major part of recent work on percolation is dedicated to tailor-made approaches for specific systems, for instance rod-like systems \cite{kyrylyuk2008continuum,drwenski2017connectedness,meyer2015percolation,nigro2013quasiuniversal,kale2015tunneling,mutiso2012simulations,jadrich2011percolation,schilling2015percolation}, which have attracted particular attention due to their use as fillers in composite materials. 

Although percolation itself is trivial in one-dimension if the connectivity range of a each individual particle remains finite \cite{schulman1983long}, resolving the distance dependence of the probability of two particles being part of the same cluster is not trivial. As this quantity exists in any dimension, an exact solution of a one-dimensional problem provides the perfect benchmark for a more general formalism.  However, to our knowledge, even in one dimension the connectivity problem has been solved exactly only for two systems: the ideal gas of non-interacting particles \cite{domb1947problem} and the system of 
impenetrable hard rods \cite{vericat1987exact,drory1997exact}. These cases were cracked in completely different ways, each tailored to the specific system. However, both solutions can be obtained straightforwardly from the same integral equation as we show below.

In this work, we derive an exact integral equation for the pair connectedness for given arbitrary pair-interactions. 
In one-dimensional systems with nearest-neighbor interactions, this integral equation requires only the pair-density as input and no approximations. For higher dimensions and long-ranged interactions, a closure relation is required, however, for an intuitively accessible quantity. We demonstrate the virtue of this perspective for the three-dimensional ideal gas.

The fundamental aim of our considerations is to provide a framework that links connectivity properties to thermal distribution functions. As approximate pair distributions are known for many interaction potentials either in analytical form or from experiments and computer simulations, an immediate link to connectivity functions is of considerable practical value. We demonstrate how thermal distribution functions can be used as input to a computational scheme that yields the corresponding connectivity properties.

We revisit established integral equations and summarize the necessary basics in 
Section \ref{sec:establishedIntegralEquations}. 
In Section \ref{sec:NNinteractions} we work out our framework for nearest-neighbor interactions 
and discuss the two analytically known test cases of non-interacting ideal and impenetrable hard-core 
particles as well as a numerical example. 
Generalizations of our approach to external fields, 
long-ranged interactions and higher dimensions, including an analysis of three-dimensional fully penetrable spheres, are discussed in Section \ref{sec:generalizations}.

\section{Established Integral Equations} \label{sec:establishedIntegralEquations}

\begin{figure}[t!]

	\centering
	\includegraphics[width=0.45 \textwidth]{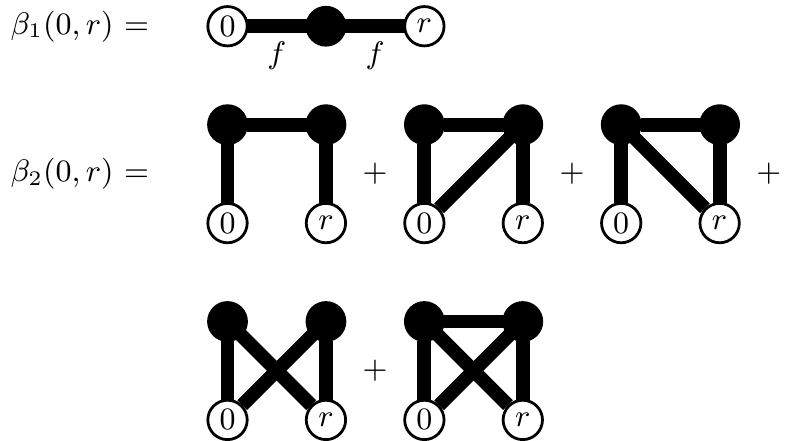}
	
	\caption{Irreducible cluster integrals 
	$\beta_1$ and $\beta_2$ 
	of the second kind 
	with $1$-circles and $f$-bonds. Black circles denote particle positions that are integrated out. 
	}
	\label{fig:betas}
\end{figure}

The quantity we focus on within this paper is the pair connectedness $P(\va{i},\va{j})$ as, for instance, defined by Coniglio \cite{coniglio1977pair} by demanding that
\begin{align}
	\rho^2 P(\va{i},\va{j}) \dd \va{i} \dd \va{j}
	\label{eq:DefP}
\end{align}
describes the absolute probability to find particles $i$ and $j$ within the corresponding volume elements $\dd\va{i}$ and $\dd\va{j}$ belonging to the same connected cluster; $\rho$ is the number density of the system. Notice that we consider particles $i$ and $j$ connected if the distance between their assigned coordinates $|\bs{r}_i-\bs{r}_j|$ is smaller than a constant threshold $d$. In that sense, the coordinates $r_i$ and $r_j$ can be regarded as centers of spherical connectivity shells of diameter $d$ so that a connection corresponds to overlapping connectivity shells. This notion of connectivity is commonly referred to as Boolean model. Definition (\ref{eq:DefP}) implies $P(\va{i},\va{j}) \leq  g(\va{i},\va{j})$, where $g$ denotes the pair-distribution function. Furthermore, we define the connection probability $p$ that particles centered at $\va{i}$ and $\va{j}$ are part of the same cluster given their existence as
\begin{align}
g(\va{i},\va{j}) p(\va{i},\va{j}) :=
P(\va{i},\va{j}) . 
\label{eq:condProb}
\end{align}
It seems natural to assume that a complete description of the 
connectivity properties requires complete information on the thermodynamic equilibrium, \textit{i.e.}, 
density distributions to arbitrary order. Acquiring this critical knowledge is commonly subsumed as solving the \textit{thermal problem}. In contrast to that, extracting the connectivity properties for given density distributions is referred to as the \textit{percolation problem}. \newline
Following reference \cite{hansen1990theory}, the pair-distribution function can be written as a  diagrammatic density expansion using the Mayer $f$-bonds 
\begin{align}
	f(\va{i},\va{j}) = \exp(-\beta V(\va{i},\va{j}))-1 \; ,
\end{align}
for an arbitrary pair potential $V$:
\begin{align}
g(i,j) = 1 + f(i,j) + (f(i,j)+1) \sum_{n=1}^\infty \rho^n \beta_n \; ,
\label{eq:gexpansion}
\end{align}
where $\beta_n$ denotes the sum over the so-called irreducible cluster integrals of the second kind of order 
$n$; numbers $i$ stand for particle positions $\va{i}$. Each diagram in $\beta_n$ contains two labeled 
white circles, $n$ black circles, and $f$-bonds between them such that there is no direct bond between the 
white circles, but if you drew an imaginary line between them, each diagram would be free of connecting circles. Removal of a connecting circle splits the diagram into two or more separate components. 
Schematic representations of $\beta_1$ and $\beta_2$ are depicted in Figure \ref{fig:betas}.

The pair connectedness can be treated with exactly the same expansion by modifying the $f$-bonds such that 
they distinguish between connected and disconnected pairs of particles. For Boolean models the Boltzmann factor $e(i,j)=f(i,j)+1$ can 
be split into a connected ($\dagger$) and a disconnected ($\ast$) part, respectively, using the Heaviside step function $\Theta$ \cite{hill1955molecular}: 
\begin{align}
	e(i,j) &=& &\Theta(d-|\va{i}-\va{j}|)e(i,j) + \Theta(|\va{i}-\va{j}|-d)e(i,j) \nonumber \\ &=:& &e^\dagger(i,j) +  e^\ast(i,j) \; . 
	\label{eq:esplit}
\end{align} 
This translates into the corresponding Mayer bonds via
\begin{align}
	f^\dagger(i,j) = e^\dagger(i,j) \qquad f^\ast(i,j) = e^\ast(i,j) - 1 \; .
	\label{eq:fsplit}
\end{align}
Applying eqn.~(\ref{eq:fsplit}) to eqn.~(\ref{eq:gexpansion}) yields an expansion of $g$ in terms of the 
\textit{connectivity bonds} $f^\dagger$ and the \textit{blocking functions} $f^\ast$. Finally, if the 
sum is restricted to diagrams that feature a connection between the white circles established purely via 
$f^\dagger$-bonds, we arrive at the density expansion of the pair connectedness function $P$: 
\begin{align}
P(i,j) &=& & g(i,j) \nonumber \\
 & =& &e(i,j) + e^\dagger(i,j) \sum_{n=1}^\infty \rho^n \beta_n 
 \quad &\mathrm{for}& \quad  |\va{i}-\va{j}| < d \; , \label{eq:p_trivial}\\
P(i,j) &=& &e^\ast(i,j) \sum_{n=1}^\infty \rho^n \beta^\dagger_n \quad &\mathrm{for}& \quad  |\va{i}-\va{j}| \geq d \; ,
\label{eq:Pexp}
\end{align}
where $\beta_n^\dagger$ contains all diagrams of $\beta_n$ which after replacing each $f$ by 
either $f^\dagger$ or $f^\ast$ feature a path of $f^\dagger$-bonds connecting the two 
white circles. 
Naturally, $P(i,j) = g(i,j)$ follows for $|\va{i}-\va{j}| < d$ from eqn.~(\ref{eq:p_trivial}). 
Eqn.~(\ref{eq:Pexp}) is the expansion that has to be reproduced by any alternative approach 
in order to be exact. \newline 
The approach to continuum percolation by Coniglio \textit{et al.}~\cite{coniglio1977pair} starts off by 
dividing the diagrams in eqn.~(\ref{eq:Pexp}) into nodal and non-nodal 
parts. The latter constitute the direct connectivity $(C^\dagger)$. Recognizing that the nodal-diagrams 
can be constructed as products of non-nodal diagrams results in the connectivity Ornstein-Zernike (OZ) 
relation 
\begin{align}
	P(i,j) = C^\dagger(i,j) + \rho \int \dd k \; C^\dagger(i,k) P(k,j) \; .
	\label{eq:COZ}
\end{align}
This approach of casting the non-nodal diagrams bears the indisputable advantage, that due to its 
structural equivalence to the standard OZ relation, the entire toolbox from liquid state theory can be 
applied.  
However, akin to liquid state theory \cite{hansen1990theory}, the direct connectivity requires approximations or assumptions as it is still an 
infinite sum of arbitrarily complicated diagrams. Hence, eqn.~(\ref{eq:COZ}) grants insight into the 
structure of $P$, but does not actually solve the problem unless the direct connectivity 
$C^\dagger$ function is known. Moreover, the percolation threshold is related to $C^\dagger(r\to\infty)$. 
Unfortunately, the closure relations employed in liquid state theory which allow for an analytic 
treatment tend to make assumptions specifically for this regime. The Percus-Yevick closure for instance 
assumes $C^\dagger(r>d) = 0$ and can thus not be expected to be accurate in predicting the 
percolation threshold. We therefore use a slightly different approach. \newline \newline
On the diagrammatic level, the thermal and the percolation problem share the same complexity. Unfortunately, with the exception of a few simple systems  
there are no exact solutions to the thermal problem available to work with. However, there are many decent approximations as well as experimental data and simulation results, which can be used as a starting point. We are hence interested in a computational scheme to construct $P$ from accessible observables like the pair-distribution function or the nearest-neighbor distribution. Both functions inherently contain the diagrams $\beta_n$ in their corresponding diagrammatic expansions which we endeavor to exploit. \newline \newline
In our deliberations, 
Volterra equations of the second kind \cite{tricomi1985integral}, \textit{i.e.}, equations of the form
\begin{align}
P(r) = I(r) + \lambda \int_{a}^r \dd x \; K(r,x) P(x) \; , 
\label{eq:VII}
\end{align}
will play a key role. 
Here, $P$ is to be determined for given functions $I$ and $K$ 
(referred to as inhomogeneity and kernel, respectively) and a real number $\lambda$. 
Notice, that the lower boundary of the integral can always be chosen as zero by supplementing the kernel 
with a respective Heaviside function $\Theta(x-a)$. 
This way, eqn.~(\ref{eq:VII}) can always be cast in a form to which Laplace transform techniques can easily be applied.
When $I$ and $K$ are specified, $P$ is obtained through simple numerics, or even analytically, if $I$ and $K$ 
are known analytically. If $I$ and $K$ are $L^2$-functions, the unique solution of eqn.~(\ref{eq:VII}) 
(except for functions that vanish almost everywhere) can be formally written as 
\begin{align}
P(r) = I(r) - \lambda \int_0^r \dd x \; H(r,x) I(x)
\label{eq:Solution}
\end{align}
where $H$ denotes the \textit{resolvent kernel} defined by
\begin{align}
H(r,x) = -\sum_{n=0}^{\infty} \lambda^n K_{n+1}(r,x)
\label{eq:Resolvent}
\end{align}
with the iterated kernels $K_n$. The latter satisfy the recurrence relation 
\begin{align}
K_1(x,y) &=& &K(x,y) \\
K_{n+1}(x,y) &=& &\int_0^x \dd z \; K(x,z) K_n(z,y) \;.
	\end{align}
This recurrence relation is essentially a formalized Picard iteration of which convergence is assured under 
the condition of $K, I \in L^2$ \cite{tricomi1985integral}, which for our purposes will always be trivially 
satisfied. In diagrammatic terms, eqn.~(\ref{eq:Resolvent}) corresponds to the sum over all chain diagrams 
with $\rho$-circles and $K$-bonds on the bounded interval $[0,r]$. The resolvent kernel satisfies the integral equation
\begin{align}
	H_\rho(r,x) = -K(r,x) + \lambda \int^r_x \dd z\; K(r,z) H_\rho(z,x) \;,
\end{align}
which depends exclusively on the integral kernel $K$. Volterra equations can thus be utilized to compute chain diagrams, which are essential to one-dimensional systems.

\section{Nearest-Neighbor Interactions}
\label{sec:NNinteractions}

In this section, we develop our approach for nearest-neighbor interactions and present exact results for pair-connectedness functions.
There are so far only two systems for which an exact analytical expression for the 
pair connectedness has been found, namely the one-dimensional ideal gas \cite{domb1947problem} and one-dimensional 
hard ``spheres'' (\textit{i.e.} impenetrable line segments) \cite{vericat1987exact,drory1997exact}. The solutions to these two cases employ vastly different techniques, however, they share the convenient property that 
interactions (if present at all) are restricted to nearest neighbors. 

Consider a system of only pairwise interacting identical particles in one dimension. In the absence of any external field, the one-particle density \footnote{$\rho^{(1)}(r)$ is defined as the grand-canonical ensemble average over all configurations featuring a particle at $r$.} $\rho^{(1)}(r)$  equals the number density $\rho$ of the interacting particles. 
We set coordinates such that one particle is fixed at the origin; without loss of generality we restrict 
our considerations to $r > 0$. The pair-distribution function $g^{(2)}(0,r) = g(|r|) = g(r)$ is a measure of the average density of particles at a distance $r$ from the origin, given that there is a particle at the origin.
In contrast to that, the pair connectedness $P$ describes the average density of particles that additionally belong to the same cluster as the particle at the origin.
Following eqn.~(\ref{eq:condProb}), $P$ can be factorized 
\begin{align}
	P(r) = p(r) g(r) \; ,
\end{align}
where $p$ is an actual probability, \textit{i.e.}, $p \in [0,1]$, the probability that two particles a distance $r$ apart belong to the same connected component.  Naturally, within the connectivity 
shell of a particle $p(r<d) \equiv 1$, hence, $P(r) = g(r)$. Beyond $d$, the connection must be 
established through at least one mediating particle within the connectivity shell of the first particle. 
The average density $\omega(\tau)$ of such particles at $\tau \in (0,d)$ can be expressed by the pair-distribution function 
\begin{align}
	\omega(\tau) = \rho g(\tau) \; . 
\end{align}
Yet, there might be more than one single particle in the connectivity range of the first particle at the origin. 
Indeed, if there were an additional particle at $\tau' \in (0,\tau)$, this in general would impact $\omega$ and 
require integrating over all possible configurations of particles in $(0,\tau)$. In order to avoid the 
difficulties connected to this problem, we instead only look for the particle closest to the origin (in positive direction), 
ruling out the existence of obstructing correlations due to intermediate particles by definition. Thus, we 
are interested in the distribution of nearest neighbors. The probability distribution $\omega'$ for finding 
such a nearest neighbor to the first particle at $\tau$, can be decomposed as 
\begin{align}
	\omega'(\tau) = \omega(\tau) \mathcal{P}((0,\tau) \; \mathrm{empty} \, | \mathrm{\; particle \, at \, \tau}) \; ,
	\label{eq:Bayes}
\end{align}
where $\mathcal{P}$ denotes a conditional probability. Torquato \textit{et al.} wrote down the reverse 
decomposition of $\omega'$ \cite{torquato1990nearest}, \textit{i.e.}, the product of the gap probability 
and the conditional existence of the particle at $\tau$. Clearly, these descriptions are equivalent through 
Bayes theorem. However, as we strive to devise a scheme that takes $g(r)$ as input, eqn.~(\ref{eq:Bayes}) bears 
the advantage that at least $\omega$ is already known. As shown in ref. \cite{torquato1990nearest}, the nearest-neighbor distribution can, in general, not be inferred from $g(r)$ alone, because an exact treatment would 
require knowledge of the entire hierarchy of density distributions. 
Yet, in one-dimensional systems with only nearest-neighbor interactions, the decomposition 
\begin{align}
	g^{(n+1)}(r_1,...,r_{n+1}) = g^{(n)}(r_1,...,r_{n})g^{(2)}(r_n,r_{n+1})
	\label{eq:Kirkwood}
\end{align}
is exact for $r_1 < r_2 < ... < r_{n+1}$ \cite{salsburg1953molecular} and with that, the entire hierarchy of higher 
order distribution functions factorizes into products of $g^{(2)}$. Thus, as long as particles only 
interact with their nearest neighbors, $g(0,r)$ contains all properties of the equilibrium system, 
for instance the nearest-neighbor distribution $\omega'(r)$, which can be constructed in the following way: \newline
The probability of finding the nearest neighbor at a position $r$ is given by the difference between $g(0,r)$ (\textit{i.e.} the probability of finding a particle at $r$ at all) and the probability of finding at least one particle in between $0$ and $r$. It might now be tempting to compute the latter probability by integrating over $g^{(3)}(0,r_1,r)$
\begin{align}
	I_1(r) := \int_0^r \dd r_1 \; g^{(3)}(0,r_1,r) \rho^{(1)}(r_1) \nonumber \\ =  \rho \int_0^r \dd r_1 \; g^{(2)}(0,r_1)g^{(2)}(r_1,r) \; .
	\label{eq:Kirk}
\end{align}
However, this expression overcounts configurations. (The reader can easily check this statement for the case of the ideal gas, where $g^{(2)}(r,r^\prime)=1 \longrightarrow I_1(r) = r\rho$, but $\omega'(r) \ne 1-r\rho$, see eqn.~(\ref{eq:omegaIdGas}).) Indeed, imagine there are exactly two particles in between 0 and $r$ placed at $r_a$ and $r_b$, respectively. The integral in eq. (\ref{eq:Kirk}) counts this configuration twice - once if $r_1 = r_a$  and again for $r_1 = r_b$ although the configuration are indistinguishable. To account for this, we need to explicitly add the configuration with particles at $0,r_a,r_b$ and $r$ with their corresponding weight given by the four point correlation function $g^{(4)}(0,r_a,r_b,r)$. Accounting for up to two mediating particles in general, we thus have to subtract
\begin{align}
I_2(r) := \rho^2 \int_0^r \dd r_1 \int_{r_1}^{r}  \dd r_2 \; g^{(4)}(0,r_1,r_2,r) \nonumber \\
=  \rho^2 \int_0^r \int_{r_1}^{r} \dd r_1 \dd  r_2 \; g^{(2)}(0,r_1) g^{(2)}(r_1,r_2) g^{(2)}(r_2,r) \; .
\end{align}

\begin{figure}[h!]
	\centering
\includegraphics[width=0.45\textwidth]{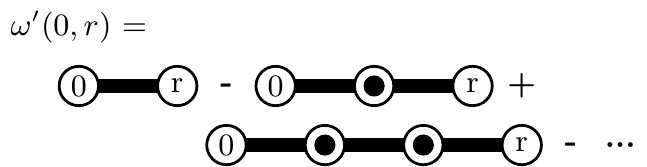}
\caption{\label{fig2} Diagrammatic expansion of the nearest-neighbor distribution $\omega'(r)$ with white 1-circles, black dotted $\rho$-circles and $g^{(2)}$-bonds. As opposed to completely black circles, the dotted ones are only integrated over the region satisfying the order condition ({\textit e.g.}~$r_1<r_2<r_3$). 
}
\label{fig:omegaexpg}
\end{figure}

However, this term now overcounts configurations with three and more particles in between $0$ and $r$. 
Continued alternating addition and subtraction of terms constructed in this way  finally yields the correct $g$-bond expansion of the nearest-neighbor distribution, see fig.~\ref{fig2}. 
\newline
The major advantage of this expansion is that it reveals the correspondence to the following integral equation
\begin{align}
	\omega'(0,r) = g(0,r) - \rho \int_0^r \dd x \; g(0,x) \omega'(x,r) \; ,
	\label{eq:NND}
\end{align}
which is a Volterra equation of the second kind, and therefore numerically dealt with. 
Thus, for systems of only nearest-neighbor interactions, the nearest-neighbor distribution can be computed 
straightforwardly from the pair-distribution function and vice versa as eqn.~(\ref{eq:NND}) is easily inverted. Perhaps, the inverted form 
\begin{align}
 g(0,r)  = \omega'(0,r) + \rho \int_0^r \dd x \; \omega'(0,x) g(x,r)  
\label{eq:NNDinv}
\end{align}
is even more intuitive as any $g(r)$ is naturally the result of a sequence of nearest neighbors. 
\newline
If we aim for two particles at $0$ and $r$ to be connected, they are either already directly connected ($r<d$), 
or a nearest neighbor of the particle located at $0$ exists within $(0,d)$ which is connected 
(through an arbitrary number of other particles) with the particle located at $r$. The corresponding integral 
equation for the pair connectedness reads 
\begin{align}
	P(0,r) =& \Theta(d-r) g(0,r) \;  \nonumber \\  +& \Theta(r-d) \rho \int_0^d \dd \tau \; \omega'(0,\tau) P(\tau,r) \; , 
	\label{eq:Main}
\end{align}
where $\Theta$ denotes the Heaviside step function. It is important to notice that this equation works for 
nearest-neighbor interactions, because the particle at the origin has no influence on the particle 
distribution beyond its nearest neighbor. Indeed, the equation suggests that the nearest neighbor can 
be chosen as a new origin from which to connect to $r$ in a shifted system of coordinates. 
Equation~(\ref{eq:Main}) is trivial for $r<d$. In absence of symmetry breaking external fields, 
the equation can be recast in the standard Volterra type for $r > d$ via:
\begin{align}
	P(r) &=& &\rho \int_{0}^{d} \dd x  \; \omega'(x)P(r-x) \nonumber \\
	 &=& &\rho \int_{r-d}^{r} \dd x \; K(r-x) P(x) \nonumber \\
	 &=& &I(r) + \rho \int_{d}^{r} \dd x \; K(r-x) P(x) , 
	 \label{eq:Main2}
\end{align}
where we introduced the kernel $K(x):=\Theta(d-x)\omega'(x)$, accounting for the fact that only 
the nearest-neighbor distribution within the initial connectivity shell has an influence on the 
connectivity properties of the system.
The inhomogeneity $I(r)$ is given by
\begin{align}
	I(r) = \rho \int_{r-d}^{d} \dd x \; \omega'(r-x) g(x)
\end{align}
for $d < r < 2d$, implementing the initial condition $P \equiv g$ for $r<d$ with $I(r \geq 2d) = 0$. 
Moreover, as presented above, 
$I$ and $K$ can both be constructed from the pair-distribution function as long as eqn.~(\ref{eq:Kirkwood}) holds:
\begin{align}
K(r) &=& & \begin{cases}
	 \rho 	g(r) -  \int_0^r \dd x \; K(r-x) g(x)  & r<d \\
	 0 & \mathrm{else}
	 \end{cases} 
\label{eq:K}
\\
I(r) &=& & \rho \int_{r-d}^{d} \dd x \; K(r-x) g(x)  .
\label{eq:I}
\end{align}

\begin{figure}[h!]
  	\centering
\includegraphics[width=0.35\textwidth]{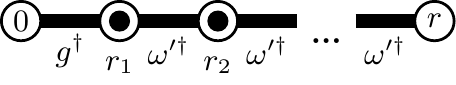}
	\caption{Connectivity carcass of a one-dimensional system. 
	 }
	\label{fig:gw}
\end{figure}

\begin{figure}[h!]
	\centering
	\includegraphics[width=0.45\textwidth]{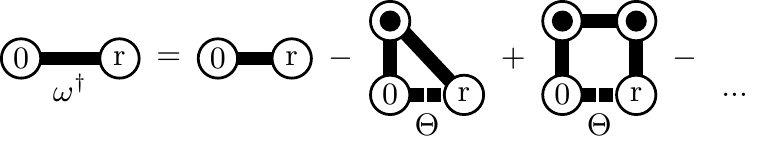}
	\caption{\label{fig4} Diagrammatic expansion of $\omega^\dagger$ in terms of $g^\dagger$-bonds (solid lines). The dashed lines represent the Heaviside bonds ensuring that the two white circles are mutually connected.}
\end{figure}

Equations (\ref{eq:Main2})-(\ref{eq:I}) suffice to recover the analytically known solutions for fully penetrable 
and impenetrable rods \cite{domb1947problem,vericat1987exact,drory1997exact} (we will show this in Subsections \ref{subsec:Id} and \ref{subsec:Imp}). While the latter has been derived in its complete form in ref.~\cite{drory1997exact} 
through a sophisticated mapping to a specific lattice model, the approach presented here is straight 
forward and, in particular, generally applicable to any kind of one-dimensional 
system with nearest-neighbor interaction. 
\newline \newline
It remains to show that eqn.~(\ref{eq:Main2}) indeed reproduces the diagrammatic representation of eqn.~(\ref{eq:gexpansion}).
Volterra equations generate ordered chain diagrams, \textit{i.e.} in our case diagrams of the structure displayed in fig.~\ref{fig:gw}
\newline
with the additional condition that $0 < r_1 < r_2 < ... < r$. We can now replace the nearest-neighbor distribution by the corresponding $g$-bond expansion, presented in fig~\ref{fig4}. 
\newline
Note that the Heaviside bond renders the additional constraint on each individual bond obsolete so that we can use $g$-bonds instead of $g^\dagger$-bonds. 
In order to relate the $g$-bond expansion to the unordered $f$-bond expansion of Coniglio \cite{coniglio1977pair}, we can identify the diagrams of a specific order in $\rho$ for both approaches. In Coniglio's expansion (see fig.~\ref{fig:betas} and eqn.~(\ref{eq:Pexp})), diagrams are naturally ordered by powers of $\rho$, \textit{i.e.} all diagrams featuring two black circles form the part of the solution that is quadratic in $\rho$. In contrast to that, the nearest-neighbor distribution as well as $g$ itself already contain contributions to arbitrary order in $\rho$. Instead, the number of dotted black circles in a diagram of our expansion defines the lower limit of nodal circles in the corresponding $f$-bond representation.

The expansions apparently coincide for $r < d$, yielding simply $g^\dagger$.
We can henceforth ignore any diagram that contains an $f$-bond between the two labeled circles. To first order in $\rho$, there remains only one diagram in Coniglio's expansion (see fig.~\ref{fig:betas} top panel). Considering
\newline
\begin{center}
		\begin{tikzpicture}
	\draw [thick] (1.5,0) circle [radius=0.2];
	\draw [fill] (2.5,0) circle [radius=0.1];
	\draw [thick] (2.5,0) circle [radius=0.2];
	\draw [thick] (3.5,0) circle [radius=0.2];
	\draw [line width = 4] (1.7,0.0) -- (2.3,0.0);
	\draw [line width = 4] (2.7,0.0) -- (3.3,0.0);
	\node at (1.5,0) {$0$};
	\node at (3.5,0) {$r$};
	\node at (2.0,-0.3) {$g^\dagger$};
	\node at (3.0,-0.3) {$\omega'^\dagger$};
	\end{tikzpicture}
\end{center}
and taking the zeroth order in $\rho$ for both bonds, we reconstruct 
\newline
\begin{center}
	\begin{tikzpicture}
	\draw [thick] (1.5,0) circle [radius=0.2];
	\draw [fill] (2.5,0) circle [radius=0.1];
	\draw [thick] (2.5,0) circle [radius=0.2];
	\draw [thick] (3.5,0) circle [radius=0.2];
	\draw [line width = 4] (1.7,0.0) -- (2.3,0.0);
    \draw [line width = 4] (2.7,0.0) -- (3.3,0.0);
	\node at (1.5,0) {$0$};
	\node at (3.5,0) {$r$};
	\node at (2.0,-0.3) {$f^\dagger$};
	\node at (3.0,-0.3) {$f^\dagger$};
	\end{tikzpicture}
\end{center}
which corresponds to the first order in Coniglio's expression except with a dotted circle instead of a proper black one. We can demonstrate that both circles are indeed equivalent by first showing more generally that the region of integration can always be reduced to the interval $[0,r]$ and further that contributions of unordered configurations cancel each other. To this end it is useful to write down $f^\dagger$ and $f^\ast$ explicitly for the specific conditions that nearest-neighbor interactions provide. Distinguishing between bonds between nearest neighbors (NN) and non-nearest neighbors as well as whether the connectivity shells of the associated particles overlap, we find 
\begin{align}
f^\dagger(r_1,r_2) = \begin{cases}
e(r_1,r_2) & \text{if} \; |r_1-r_2| \leq d \; \text{and NN} \\
1 & \text{if} \; |r_1-r_2| \leq d \; \text{and not NN} \\
0 & \text{if} \; |r_1-r_2| > d  \; \text{and NN} \\
0  & \text{if} \; |r_1-r_2| > d \; \text{and not NN} 
\end{cases}
\end{align}
\begin{align}
f^\ast(r_1,r_2) = \begin{cases}
-1 & \text{if} \; |r_1-r_2| \leq d \; \text{and NN} \\
-1 & \text{if} \; |r_1-r_2| \leq d \; \text{and not NN} \\
e(r_1,r_2)-1  & \text{if} \; |r_1-r_2| > d \; \text{and NN} \\
0  & \text{if} \; |r_1-r_2| > d \; \text{and not NN}  \; .
\end{cases}
\label{eq:fast}
\end{align}
The integral $\beta^\dagger_1$, \textit{i.e.}
\begin{center}
	\begin{tikzpicture}
	\draw [thick] (1.5,0) circle [radius=0.2];
	\draw [fill] (2.5,0) circle [radius=0.1];
	\draw [fill] (2.5,0) circle [radius=0.2];
	\draw [thick] (3.5,0) circle [radius=0.2];
	\draw [line width = 4] (1.7,0.0) -- (2.3,0.0);
	\draw [line width = 4] (2.7,0.0) -- (3.3,0.0);
	\node at (1.5,0) {$0$};
	\node at (3.5,0) {$r$};
	\node at (2.0,-0.3) {$f^\dagger$};
	\node at (3.0,-0.3) {$f^\dagger$};
	\end{tikzpicture}
\end{center}
does not contribute to $P$ if $r>d$, as the preceding $e^\ast$-bond vanishes in that case. Thus, if the position $r_1$ of the intermediate particle in the diagram above is not located within $[0,r]$, either $|r-r_1|$ or $|r_1-0|$ is larger than $d$ thanks to the one-dimensional nature of the system. This, however, implies that one of the $f^\dagger$-bonds and hence the entire diagram vanishes. This argument can be applied to any diagram of the expansion but in a slightly adapted form.  Consider the diagram
\begin{center}
	\begin{tikzpicture}
	\draw [fill] (3.5,-3) circle [radius=0.2];
	\draw [fill] (4.5,-3) circle [radius=0.2];
	\draw [thick] (3.5,-4) circle [radius=0.2];
	\draw [thick] (4.5,-4) circle [radius=0.2];
	\draw node at (3.5,-4) {0};
	\draw node at (4.5,-4) {r};
	\draw node at (3.5,-2.65) {$r_1$};
	\draw node at (4.5,-2.65) {$r_2$};

	\draw node at (4.1,-2.7) {$f^\dagger$};
	\draw node at (3.2,-3.5) {$f^\dagger$};
	\draw node at (4.8,-3.5) {$f^\dagger$};
	
	\draw [line width = 4] (3.7,-3) -- (4.5,-3.0);
	\draw [line width = 4] (3.5,-3.8) -- (3.5,-3.2);
	\draw [line width = 4] (4.5,-3.8) -- (4.5,-3.2);
	\end{tikzpicture}
\end{center}
for $r>d$. The integrand of the integral corresponding to this diagram does not necessarily vanish if one of the mediating particles lies outside of $[0,r]$, for example $r_1 < 0$ while $r_2 \in [0,r]$. That means that the bond between $r_1$ and $r_2$ connects non-nearest neighbors, so that the associated $f^\dagger$-bond becomes unity. But then the expansion also contains the diagram
\begin{center}
	\begin{tikzpicture}
	\draw [fill] (3.5,-3) circle [radius=0.2];
	\draw [fill] (4.5,-3) circle [radius=0.2];
	\draw [thick] (3.5,-4) circle [radius=0.2];
	\draw [thick] (4.5,-4) circle [radius=0.2];
	\draw node at (3.5,-4) {0};
	\draw node at (4.5,-4) {r};
		\draw node at (3.5,-2.65) {$r_1$};
	\draw node at (4.5,-2.65) {$r_2$};
	\draw node at (4.1,-2.7) {$f^\dagger$};
	\draw node at (3.2,-3.5) {$f^\dagger$};
	\draw node at (4.8,-3.5) {$f^\dagger$};

	\draw [line width = 4] (3.7,-3) -- (4.5,-3.0);
	\draw [line width = 4] (3.5,-3.8) -- (3.5,-3.2);
	\draw [line width = 4] (4.5,-3.8) -- (4.5,-3.2);
	\draw [line width = 4, densely dotted] (3.67,-3.85) -- (4.5,-3.0);
	\end{tikzpicture}
\end{center}
where the dashed line represents an $f^\ast$-bond. Since $|r_2-r_1| < d$ (otherwise $f^\dagger(r_1,r_2) = 0$ anyway) also $|r_2-0| < d$ so that $f^\ast(0,r_2) = -1$. Therefore, the two integrals considered differ only by sign and thus annihilate each other. For any configuration in which the connection of $f^\dagger$-bonds between the white circles features particles not within $[0,r]$, we can repeat this procedure. We link two particles, which the ``outlying particle'' shares $f^\dagger$-bonds with, by an $f^\ast$-bond and drop that configuration from the expansion.

However, there is one exception if the particles one would like to link by an $f^\ast$-bond are already linked by an $f^\dagger$-bond. In this case, the path through the outlying particle is obsolete. One can replace the $f^\dagger$-bonds it is connected to by $f^\ast$-bonds to obtain different diagrams of the same expansion. One might therefore replace the obsolete bonds immediately by the corresponding $f$-bonds of which there is at least one that connects non-nearest neighbors and hence vanishes. This way it becomes apparent on the diagrammatic level, that the configuration of particles outside of the interval we are interested to bridge, does not influence the connectivity properties within that interval as long as we deal with nearest-neighbor interactions.

Using the same line of reasoning one can show that all configurations that contain an $f^\dagger$-bond between non-nearest neighbors will be canceled. Yet, this argument does not directly restrict the $f^\ast$-bonds. It should be noted that there cannot be $f^\ast$-bonds between nearest neighbors, because a continuous path of $f^\dagger$-bonds between the white circles would then require at least one $f^\dagger$-bond between non-nearest neighbors, which we ruled out. Moreover,  $f^\ast(r_1,r_2)$ does also vanish for non-nearest neighbors if $|r_1-r_2|>d$. Therefore, all appearing bonds are in fact short ranged.
\newline
\newline
In summary, for nearest-neighbor interactions in one dimension, the expansion by Coniglio, eqn.~(\ref{eq:Pexp}), contains all diagrams with $f^\dagger$-bonds only between nearest neighbors, $f^\ast$-bonds only between non-nearest neighbors, and an $e^\ast$-bond between the white circles which are free of connecting circles. 
\newline
\newline
Now we take a second look at eqn.~(\ref{eq:fast}) and notice 
\begin{align}
f^\ast(r_1,r_2) = -\Theta(d-|r_1-r_2|) 
\end{align}
for non-nearest neighbor interaction.
That at hand we can rewrite our $g^\dagger$-bond expansion of fig.~\ref{fig4} by replacing the $\Theta$-bonds by $-f^\ast$ bonds
 	\begin{tikzpicture}
 	\draw [thick] (-3.25,0.37) circle [radius=0.2];
 	\node at (-3.25,0.37) {0};
 	\draw [thick] (-2.25,0.37) circle [radius=0.2];
 	\node at (-2.25,0.37) {r};
 	\draw [line width=4] (-3.05,0.37) -- (-2.45,0.37);
 	\node at (-2.75,0.07) {$\omega^\dagger$ };

 	\node at (-1.75,0.37) {$=$ };
 	
 	\draw [thick] (-1.25,0.37) circle [radius=0.2];
 	\node at (-1.25,0.37) {0};
 	\draw [thick] (-0.5,0.37) circle [radius=0.2];
 	\node at (-0.5,0.37) {r};
 	\draw [line width=4] (-1.05,0.37) -- (-0.7,0.37);
 	\node at (-0,0.37) {$+$ };
 	
 	\draw [thick] (0.5,0) circle [radius=0.2];
 	\node at (0.5,0) {0};
 	\draw [thick] (0.5,0.75) circle [radius=0.2];
 	\draw [fill] (0.5,0.75) circle [radius=0.1];
 	\draw [thick] (1.25,0.0) circle [radius=0.2];
 	\node at (1.25,0.0) {r};

 	\draw [line width=4] (0.5,0.2) -- (0.5,0.55);
 	\draw [line width=4,densely dotted] (0.7,0.0) -- (1.05,0.0);
 	\node at (0.875,-0.3) {$f^\ast$ };
 	\draw [line width=4] (0.65,0.63) -- (1.11,0.14);
 	
 	\node at (1.75,0.37) {$+$ };
 	
 	\draw [thick] (2.25,0) circle [radius=0.2];
 	\node at (2.25,0.0) {0};
 	\draw [thick] (2.25,0.75) circle [radius=0.2];
 	\draw [fill] (2.25,0.75) circle [radius=0.1];
 	\draw [thick] (3,0.0) circle [radius=0.2];
 	\node at (3.0,0.0) {r};
 	\draw [thick] (3,0.75) circle [radius=0.2];
 	\draw [fill] (3,0.75) circle [radius=0.1];
 	
 	\draw [line width=4] (2.25,0.2) -- (2.25,0.55);
 	\draw [line width=4, densely dotted] (2.45,0.0) -- (2.8,0.0);
 	 	\draw [line width=4, densely dotted] (2.4,0.15) -- (2.85,0.6);
 	\node at (2.625,-0.3) {$f^\ast$ };
 	\draw [line width=4] (3.0,0.55) -- (3,0.2);
 	\draw [line width=4] (2.45,0.75) -- (2.8,0.75);
 	\draw [line width=4,densely dotted] (0.7,0.0) -- (1.05,0.0);

 	\node at (3.5,0.37) {$+$ };
 	\node at (4.0,0.0) {$... \quad ,$ };
 	\end{tikzpicture} 
thereby eliminating the alternating sign. Then we can insert the expansion of $g^\dagger$ and exploit the same arguments as before to find that both expansions can indeed be brought in perfect unison. 

However, much more straightforwardly, we can simply put the equation up to the practical test by applying it to problems for which the exact solution is known or at least easily obtained through simulations.

\subsection{Fully Penetrable Rods}
\label{subsec:Id}
\begin{figure}
	\centering
	\includegraphics[width=0.45 \textwidth]{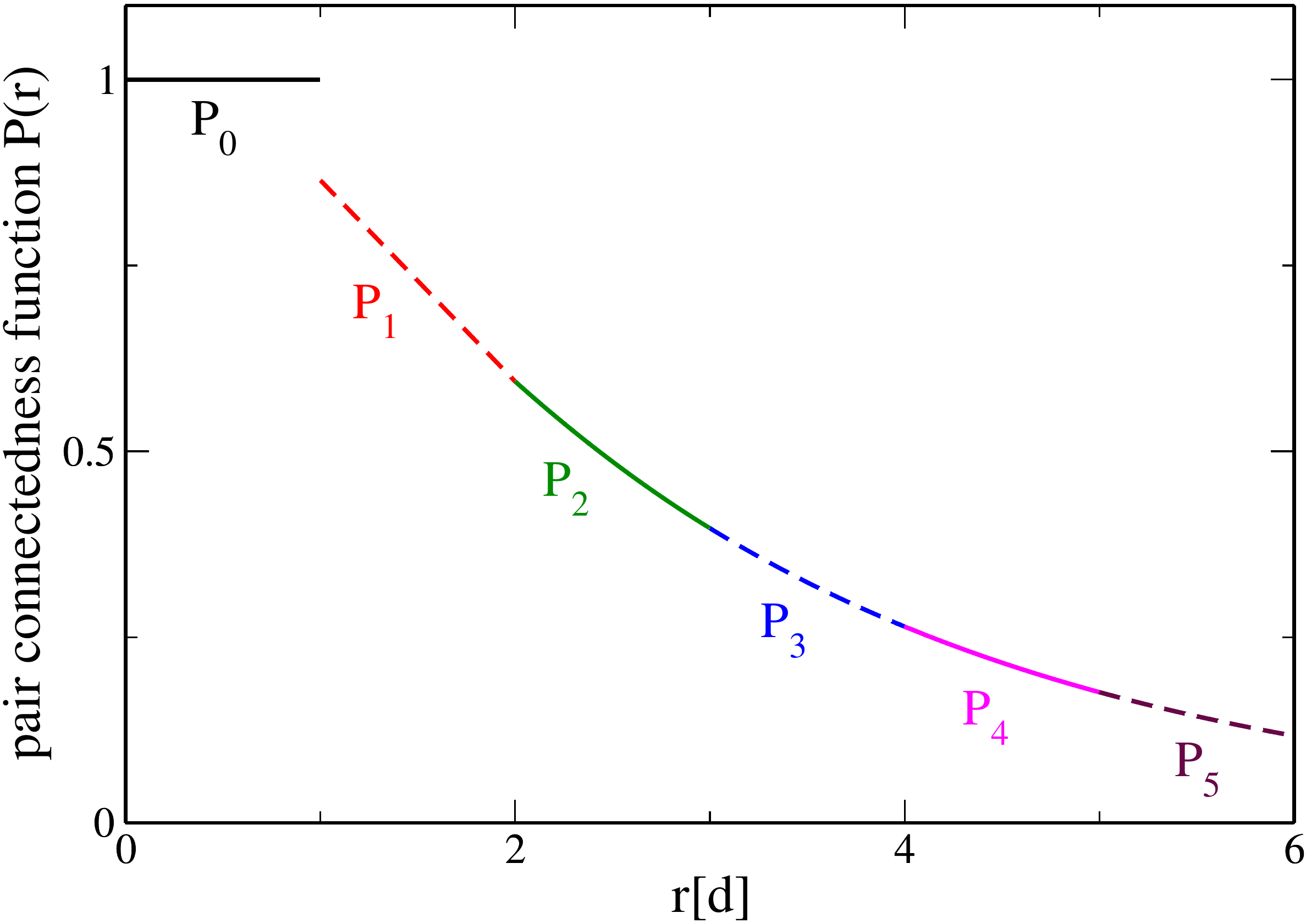}
	\caption{\label{fig5} Pair connectedness for one-dimensional fully penetrable rods of number density $\rho = 2.0$  - line styles (colors online) indicate different orders in the recurrence relation eqn.~(\ref{eq:rec22}).}
\end{figure}

A one-dimensional ideal gas, \textit{i.e.}, non-interacting fully penetrable connectivity shells, 
can be solved purely by stochastic tools. However, since the presented framework is straightforward 
to apply, the ideal gas makes up for a nice test-case system. The integral kernel follows immediately 
from eqn.~(\ref{eq:K}): 
\begin{align}
	\omega'(r) &=& &\left[1 - \rho \int_0^r \dd x \; \omega'(x) \right]  \\
  \implies \quad \omega'(r) &=&  &\exp(-\rho r) \; ,
    \label{eq:omegaIdGas}
\end{align}
recovering the well-known exponential distribution of `gap lengths' \cite{zernike1927beugung}. 
With $\omega'(r)$ known, the inhomogeneity $I$ is readily obtained using eqn.~(\ref{eq:I}): 
\begin{align}
	I(r) &=&  &\rho \int_{r-d}^d \dd x \;  \exp(-\rho x) \nonumber \\ 
	&=&  &\Theta(2d-r) \left[ e^{-\rho (r-d)} - e^{-d \rho} \right] . 
\end{align}
The integral equation for the pair connectedness function of the one-dimensional ideal gas thus reads 
\begin{align}
	P(r) =& &\Theta(2d-r) \left[ e^{-\rho (r-d)} - e^{-d \rho} \right] \nonumber \\ +& & \rho \int_{d}^{r} \dd x  \; e^{-\rho (r-x)}\Theta(d-(r-x)) P(x) . 
	\label{eq:FPR_Main}
\end{align}
The analytical solution via the resolvent 
kernel $H$ of eqn.~(\ref{eq:Resolvent}) is intricate as the ensuing Heaviside integrals are not 
straightforward to compute. Yet, the Heaviside functions can be eliminated by restricting the integral 
equation to intervals $[nd,(n+1)d]$ and, progressively, solving it for $n\in \mathbb{N}$. 
For $n=1$, eqn.~(\ref{eq:FPR_Main}) is simplified to 
\begin{align}
P_1(r) =& & \left[ e^{-\rho (r-d)} - e^{-d \rho} \right] \nonumber + \rho \int_{d}^{r} \dd x  \; e^{-\rho (r-x)} P_1(x) \; ,
\end{align}
which is solved by the linear function
\begin{align}
	P_1(r)= 1- e^{-\rho d}(\rho r- \rho d+1) . 
\end{align}
In general, the solution can be found by assuming a polynomial of $n$-th degree and comparing
the coefficients of all but the leading order in $r$ as well as the coefficient of the $e^{-\rho r}$ term. The equation for the coefficient of leading order is always trivially satisfied. Thus there are $n+1$ equations for $n+1$ coefficients, granting the unique solution. The procedure can be generalized to yield the 
solution for all $n$ in form of the following recurrence relation: 
\begin{align}
	&P_n(r) &=& \; P_{n-1}(r) - \frac{\rho^{n-1}}{n!}e^{-\rho n d}(n d - r)^{n-1}(\rho r - \rho n d + n) \nonumber \; \\
	&P_0 &\equiv& \; 1 \; .
	\label{eq:rec22}
\end{align} 
Once cast in a closed form, eqn.~(\ref{eq:rec22}) recovers the known solution obtained before by Domb and 
others \cite{domb1947problem,torquato2013random}. 
The solution is shown in fig.~\ref{fig5}. 

\subsection{Impenetrable Rods}
\label{subsec:Imp}
\begin{figure}
	\centering
	\includegraphics[width=0.45 \textwidth]{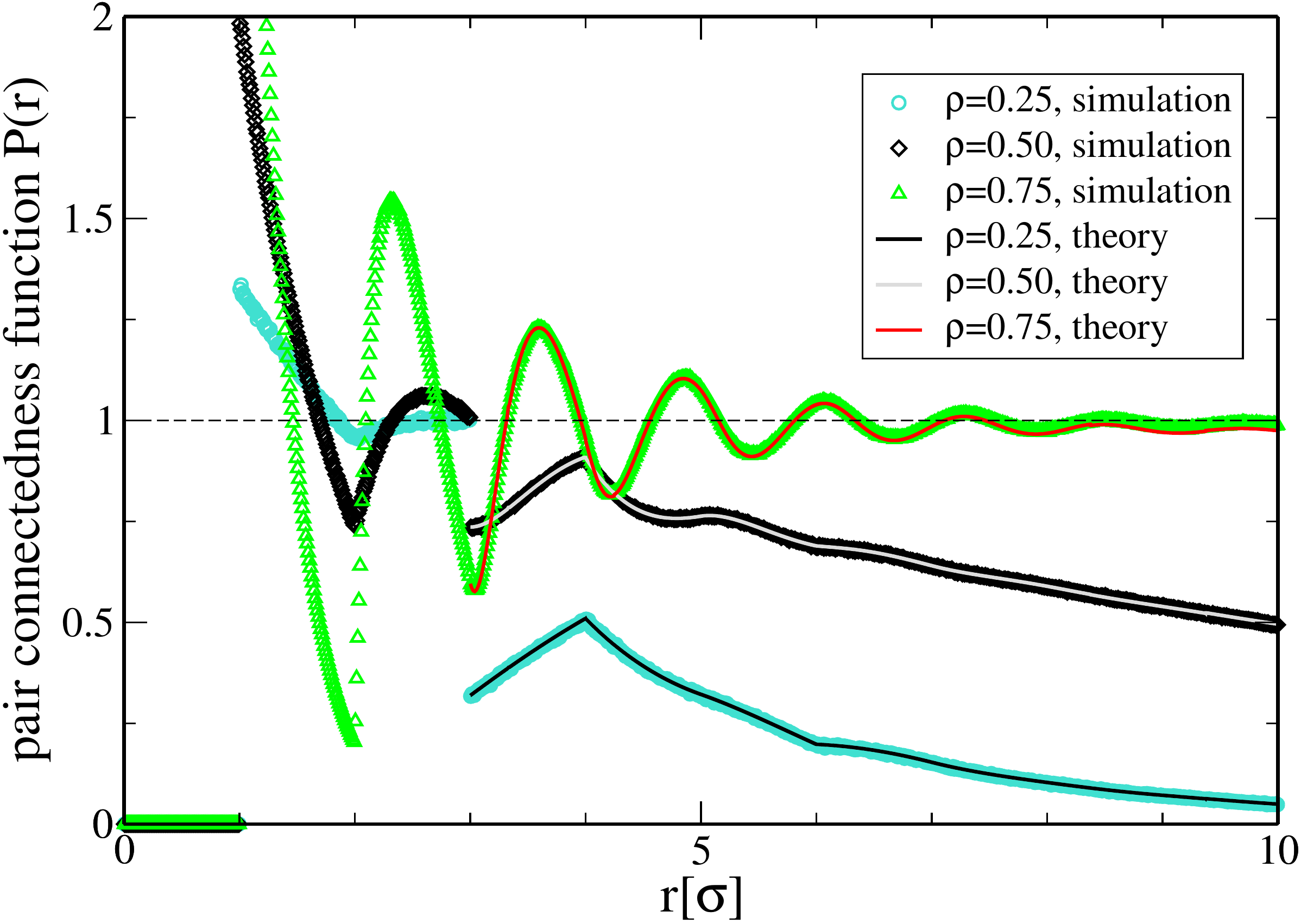}
	\caption{\label{fig6} Pair connectedness for one-dimensional impenetrable rods for different number densities - solid lines correspond to eq.~(\ref{eq:IRG:P}) - symbols denote the corresponding simulation result. }
      \end{figure}

One of the few non-trivial systems
for which the pair-distribution function has been found exactly in an analytic closed form, 
is the system of one-dimensional impenetrable identical hard rods \cite{vericat1987exact,drory1997exact}. 
For instance from classical density functional theory it is known that the corresponding pair-distribution function reads \cite{percus1976equilibrium} 
\begin{align}
	g(r) = \sum_{k=0}^\infty \Theta(r-(k+1)\sigma) \frac{\rho^k (r-(k+1)\sigma)^k}{k!(1-\rho \sigma)^{k+1}}e^{\frac{-\rho(r-(k+1)\sigma)}{1-\rho \sigma}} , 
	\label{eq:IRG}
\end{align}
where $\sigma$ denotes the length of the rods. 
The nearest-neighbor distribution can be computed from eqn.~(\ref{eq:NND}). 
However, the distance distribution to a nearest neighbor is also equivalent to the gap length distribution. 
This, in turn, can be understood as randomly (\textit{i.e.} in a uniformly distributed manner) placing $N$ points on a line 
of a length that corresponds to the free volume. The corresponding probability distribution has already 
been formulated by Zernike \cite{zernike1927beugung} in the form 
\begin{align}
	\omega'(x) = \frac{e^{-\frac{\rho(x-\sigma)}{1-\rho \sigma}}}{1-\rho \sigma} \Theta(x-\sigma) \; ,
	\label{eq:IRNND}
\end{align}
which is simply the zeroth order term in eqn.~(\ref{eq:IRG}). Hence the integral kernel is yet again an exponential, which implies that the resolvent kernel is similar in structure as well. Indeed, the same procedure that worked for the ideal gas also works for impenetrable spheres. As a result, the solution previously reported by Drory \cite{drory1997exact} (and partially before also in ref.~\cite{vericat1987exact}) as 
\begin{align}
	P(r)=\frac{1}{\eta} \sum_{k=0}^{\infty} \sum_{j=0}^{k} \frac{(-1)^j k!}{j!(k-j)!(k-1)!}\left(\frac{\eta}{1-\eta}\right)^{k} \nonumber  \\ \times \left[ r+j-k+jd \right]^{k-1}
	\Theta(r+j-k-jd)e^{\frac{-\eta(r-k)}{1-\eta}}
	\label{eq:IRG:P}
\end{align}
is found to be the unique solution to eqn.~(\ref{eq:Main}), with $g$ and $\omega$ defined by eqn.~(\ref{eq:IRG}) 
and eqn.~(\ref{eq:IRNND}), respectively. 
The agreement of the theory with simulations is shown in fig.~\ref{fig6}. 

Note, that we used the known solution for the thermal problem for simplicity. This is  not required here, as the thermal problem can also be mapped onto a Volterra equation. For impenetrable rods $g$ can be obtained by solving
\begin{align}
	1-g(r) =& \Theta(\sigma-|r|) - \nonumber \\ -\frac{\rho}{1-\sigma \rho}\Theta(|r|-\sigma)\int_0^r &\dd x \; \Theta(\sigma-|r-x|)  (1-g(x)) \; . 
\end{align}
Note further, that on the diagrammatic level, the hard core repulsion is not a nearest-neighbor interaction but rather short ranged, as the $e$-bond between non-nearest neighbors is not necessarily unity. Integral equation (\ref{eq:Main2}) remains perfectly valid, but the line of reasoning in the comparison to the more general expansion eqn.~(\ref{eq:Pexp}) has to be slightly modified.  
      
\subsection{Numerical Examples}
\begin{figure}[h!]
	\centering
	\includegraphics[width=0.45 \textwidth]{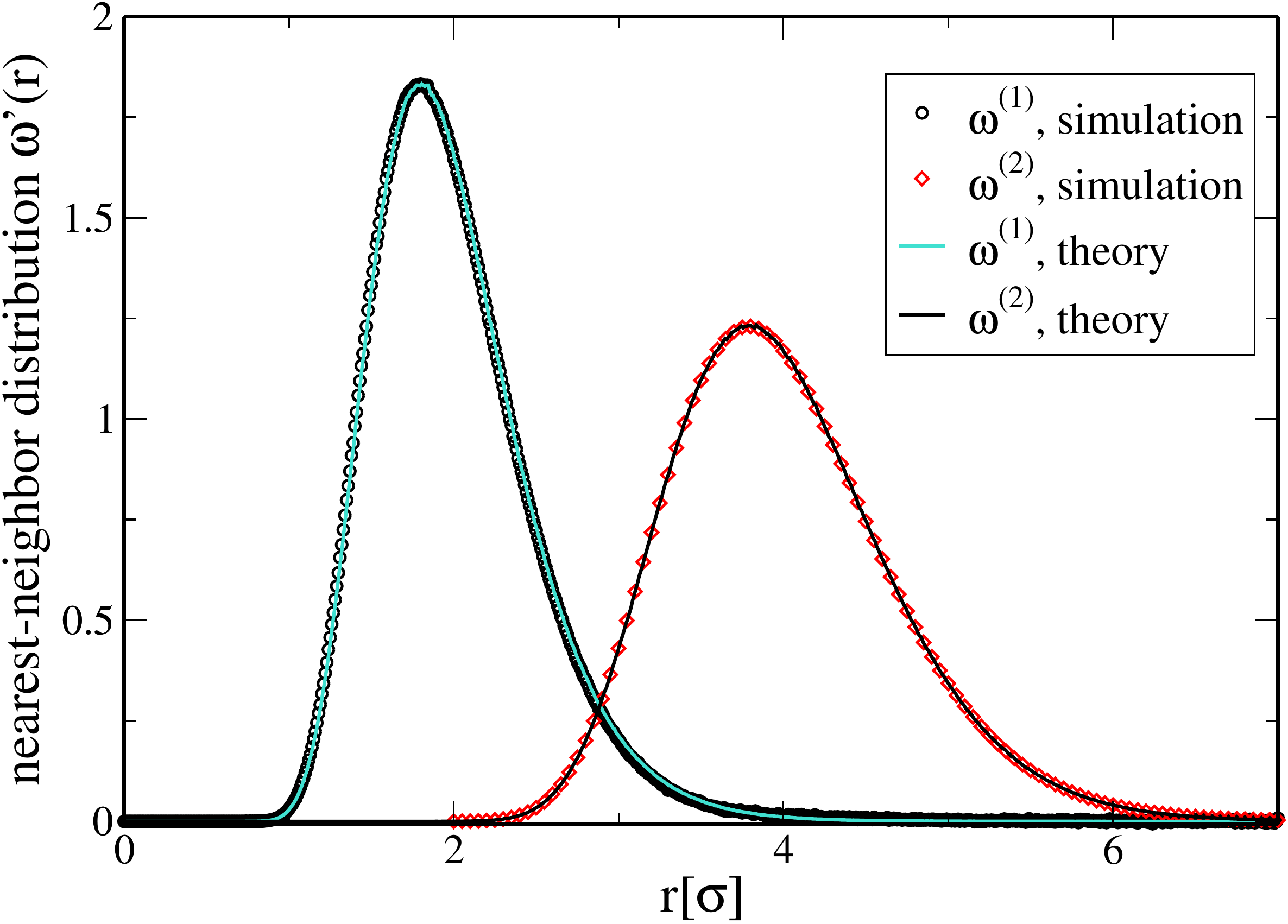} 
	\caption{Nearest- and Next-Nearest-Neighbor distribution for $V_{10}$ via simulation (symbols) and eqn.~(\ref{eq:NND}) (lines).
	\label{fig:NND} }
\end{figure}

\begin{figure}[h!]
	\includegraphics[width=0.49 \textwidth]{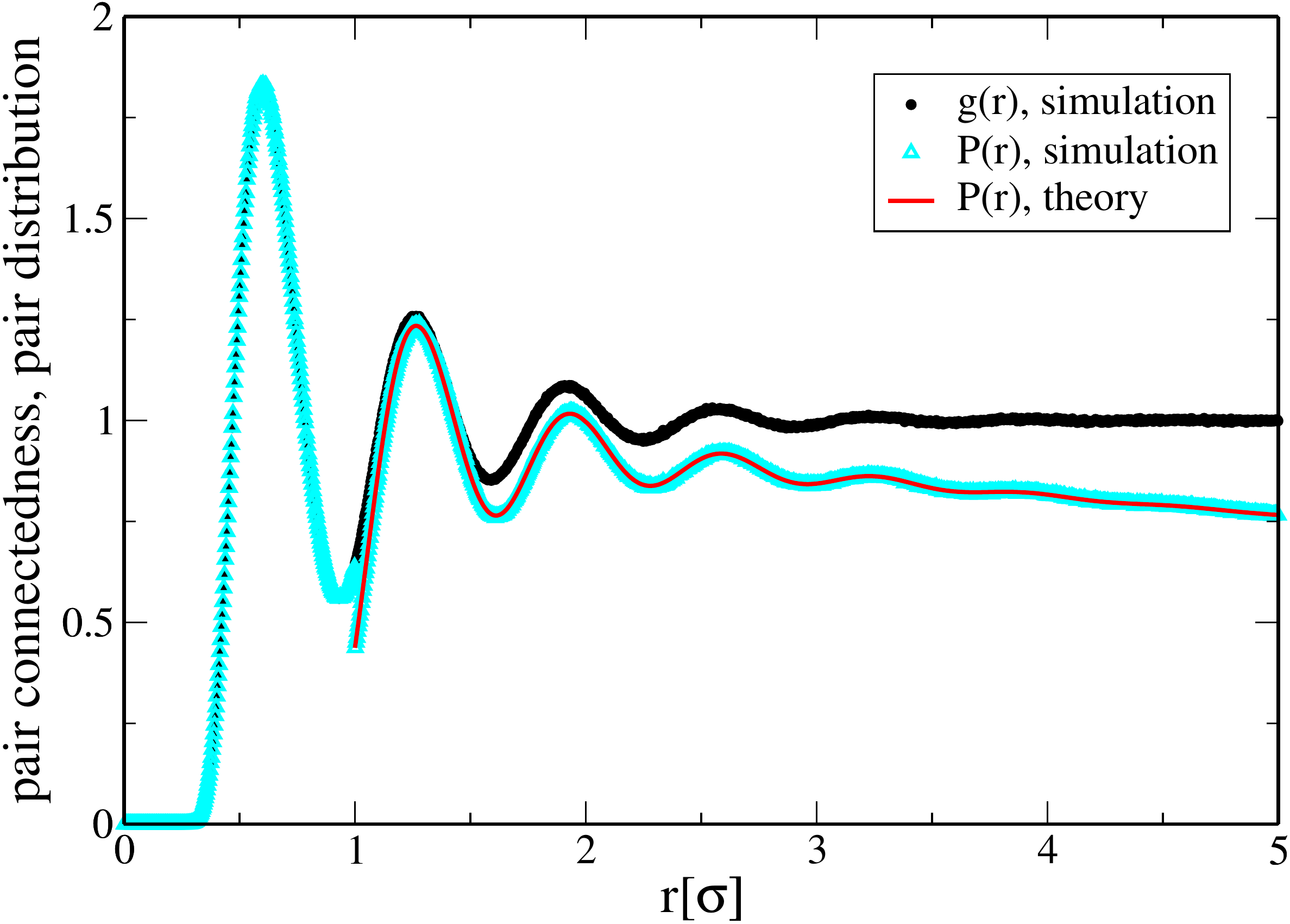} 
	\caption{Comparison between the solution of eqn.~(\ref{eq:Main2}) and simulation results for the pair connectedness induced by $V_{10}$. 
	\label{fig:IS}}
\end{figure}

\begin{figure}[h!]
	\includegraphics[width=0.49 \textwidth]{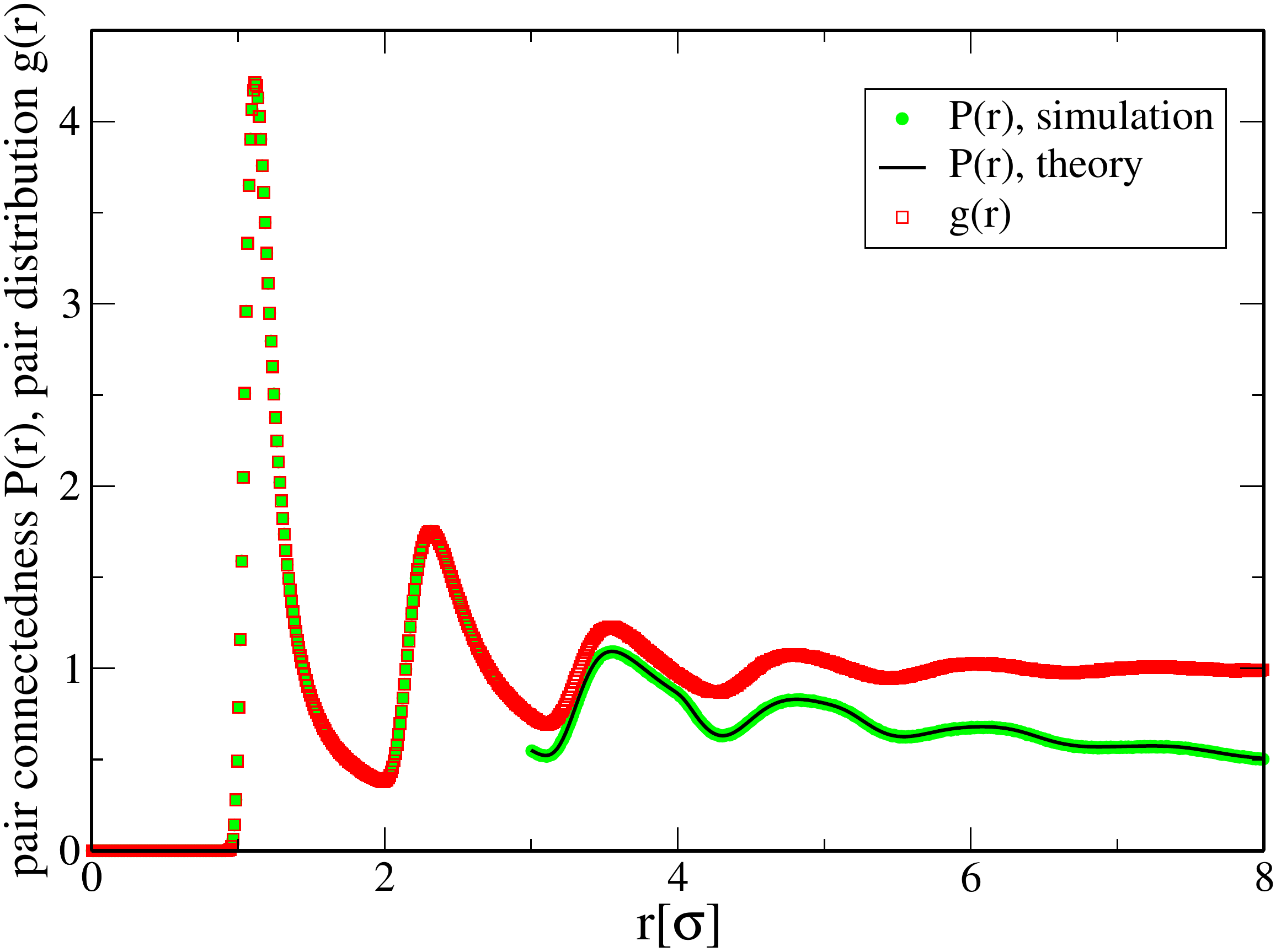} 
	\caption{Pair connectedness for  the one-dimensional Lennard-Jones fluid of number density $\rho = 0.5$, Lennard-Jones parameters $\varepsilon_{\rm LJ} = 2$, $d = 3\sigma_{\rm LJ}$, with a cutoff at separations exceeding $10\sigma_{\rm LJ}$. 
	\label{fig:LJ}}
\end{figure}

\begin{figure}
	\includegraphics[width=0.49 \textwidth]{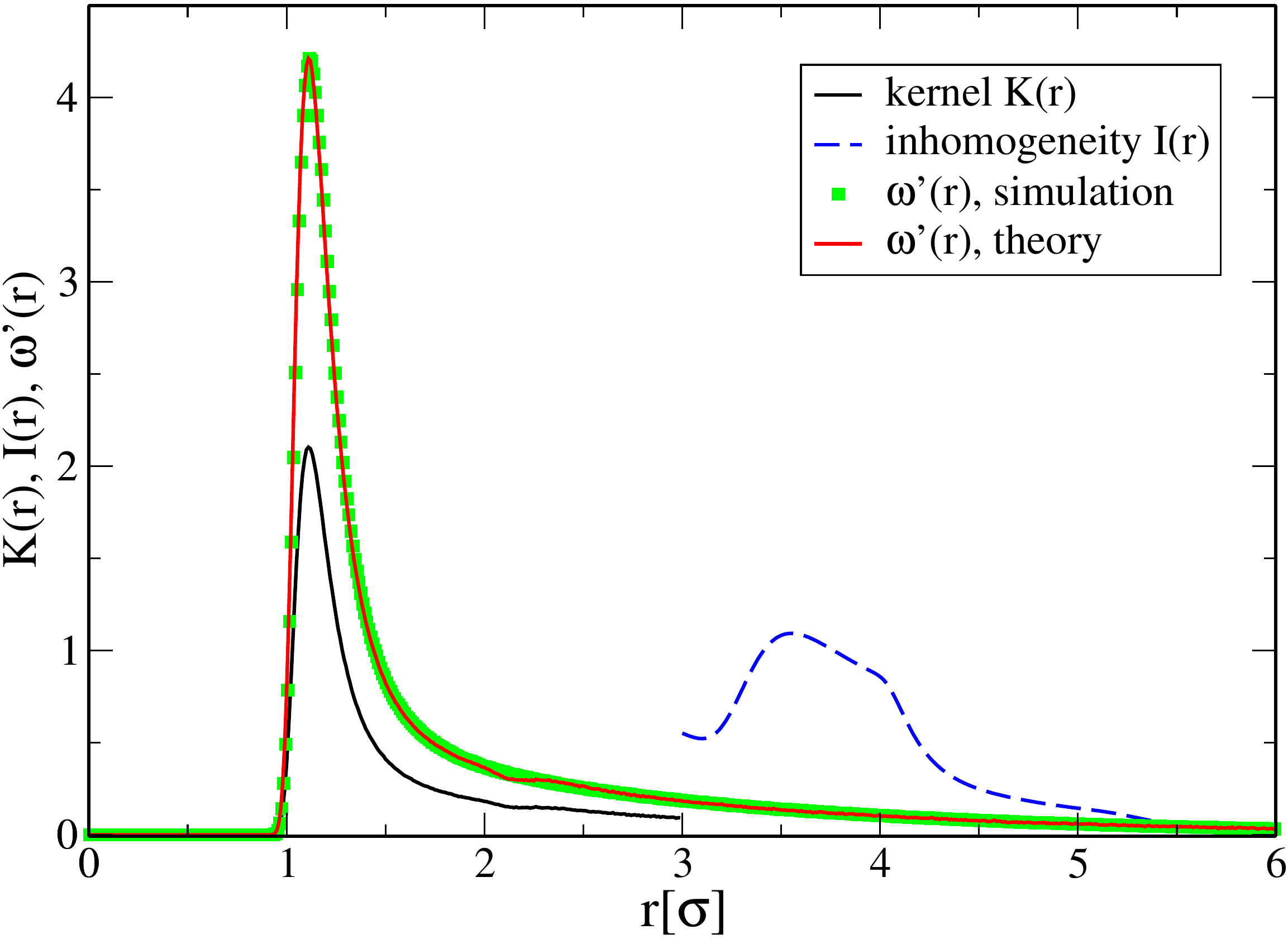} 
	\caption{Auxiliary functions for the Lennard-Jones fluid for the same parameters as in fig. \ref{fig:LJ}.}
		\label{fig:LJ2}
\end{figure}

So far we were able to reproduce known results straightforwardly because the associated thermal distribution functions were available in a closed analytical form. However, one major virtue of the proposed scheme lies in the fact that it does not require analytic input to work. We can simply sample the pair-distribution function in the first connectivity shell for an arbitrary nearest-neighbor interaction and compute the pair connectedness. Thus we can, for instance, perform a Monte Carlo simulation to extract $g(r)$, numerically solve the ensuing Volterra equation on an equidistant grid to obtain predictions for the pair connectedness and compare them to the simulations. 
To demonstrate this, we consider the purely repulsive pair potential 
\begin{align}
	V_\varepsilon(r_i,r_j) = (\delta_{i+1,j} + \delta_{i-1,j}) \varepsilon \frac{ \sigma^2}{|r_i-r_j|^2} . 
\end{align}
Solving eqn.~(\ref{eq:NND}) for this interaction results in the nearest-neighbor distribution depicted in figure \ref{fig:NND}. As the potential acts only on nearest neighbors, the next-nearest-neighbor distribution $\omega^{(2)}$ is simply the convolution of $\omega^{(1)}:= \omega'$ with itself
\begin{align}
	\omega^{(2)}(0,r) = \int_0^r \dd x \; \omega^{(1)}(0,x) \;  \omega^{(1)}(x,r) \; ,
	\label{eq:NNND}
\end{align}
which is also shown in fig.~\ref{fig:NND}. Once the hierarchy of nearest-neighbor distributions and therefore the kernel of our integral equation is known, the problem becomes trivial. 
We solve eqn.~(\ref{eq:Main2}) numerically yielding the solid line in fig.~\ref{fig:IS} which as expected is in perfect agreement with the pair connectedness determined by simulations. Notice, that the process is even invertible, \textit{i.e.}~from the pair connectedness the kernel can be reconstructed, yielding the nearest-neighbor distribution which will give you the radial distribution function. That means, for one-dimensional nearest-neighbor interacting systems the pair-distribution functions contains the same information as the pair connectedness. \newline
 \newline
For short-ranged but not necessarily nearest-neighbor interactions we cannot expect eqn.~(\ref{eq:Main2}) to be exact. However, it serves as a good approximation if the interaction energy resulting from beyond nearest neighbors is small, \textit{i.e.}~if the potential decays sufficiently fast. As an example, for particles interacting through the Lennard-Jones potential
\begin{align}
	V_{\rm LJ}(r_i,r_j) =  4 \varepsilon_{\rm LJ} \left[ \left(\frac{r}{\sigma_{\rm LJ}} \right)^{-12}  - \left(\frac{r}{\sigma_{\rm LJ}} \right)^{-6} \right]
\end{align}
 theory and simulation results for the pair connectedness cannot be distinguished by eye (see fig.~\ref{fig:LJ}). 
\newline
If you turn your attention to the nearest-neighbor distribution in fig.~\ref{fig:LJ2}) at large $r$, a slight discrepancy is visible between the simulation data and the theory. Yet, the deviation appears in a regime where the nearest-neighbor distribution is already small whereas the main peak is properly depicted. Moreover, the inhomogeneity 
\begin{align}
I(r) = \rho \int_{r-d}^{d} \dd x \; \omega'(r-x) g(x)\; ,
\nonumber
\end{align}
as the above convolution, weighs the nearest-neighbor distribution around $d$ with the short range $g$ which for any close to hard-core-interaction should be extremely small. The inaccuracy therefore hardly propagates. 
While this level of agreement seems to be a lucky coincidence, the following section illustrates how long-ranged interactions can be treated in a more systematic way. (It shall be noted though that the solution of Volterra equations is typically stable against noise in the input.) 
\section{Generalizations}
\label{sec:generalizations}
Arguably, one-dimensional classical systems with nearest-neighbor interactions do not occur in real life on a regular basis. We therefore strive to generalize the depicted procedure to more realistic conditions and discuss the impact of the added complexity.
\subsection{External Fields}
If we stay in one dimension for a start, 
an external field $\phi$ destroys the homogeneity of the system such that the single-particle density $\rho^{(1)}$ is not a constant throughout the system anymore. With that, all distribution functions become explicitly dependent on the positions they are evaluated for. However, eqn.~(\ref{eq:Main}) formally already accounted for a potential dependence on two points in space. Thus, the only modification required is to replace the number density by the space-dependent single-particle density:
\begin{align}
	&P_\phi(r_1,r_2) = \Theta(d-|r_1-r_2|) g_\phi(r_1,r_2) \;  \nonumber \\  &+ \Theta(|r_1-r_2|-d)  \int_{r_1}^{r_1+d} \dd \tau \; \omega'_\phi(r_1,\tau) \rho^{(1)}_\phi(\tau) P_\phi(\tau,r_2) \; ,
\end{align}
assuming $r_1 < r_2$ for simplicity.
Supplemented by the \textit{initial condition}
\begin{align}
	P(r_1,r_2) = g(r_1,r_2) \quad  \mathrm{for} \quad |r_1-r_2| < d \; ,
\end{align} 
this equation is also exact for nearest neighbor-interactions, however, it requires the density profile $\rho^{(1)}(\tau) $ as additional input. Moreover, $g, \omega'$ and most importantly $P$ depend on two points in space, such that the equation becomes technically a two-dimensional Volterra equation which is numerically more challenging than its one-dimensional counterpart. The important observation is that on tagging a particle the system still entirely splits, in that there is no coupling between left and right hand side of the particle, respectively. The external field adds a local weight to the integral kernel but that is all there is to it.

Cast in the standard form, the equations that need to be solved read
\begin{align}
	&P_\phi(r_1,r_2) &= & \; I_\phi(r_1,r_2) \;  \nonumber+ \\ &&&\int_{r_1}^{r_1+d} \dd x \; \rho^{(1)}_\phi(x) K_\phi(r_1,x)\rho^{(1)}_\phi P_\phi(x,r_2) \nonumber \\ 
	&I_\phi(r_1,r_2) &= & \; \int_{r_2-d}^{r_1+d} \dd x \; \rho^{(1)}_\phi(x) K_\phi(r_1,x) g_\phi(x,r_2) \nonumber \\
	&K_\phi(r_1,r_2) &= & \; \omega'_\phi(r_1,r_2)\Theta(d-|r_2-r_1|) \; \nonumber \\
	&\omega'_\phi(r_1,r_2) &= & \; g_\phi(r_1,r_2) -  \nonumber\\
	&&&\int_{r_1}^{r_2} \dd x \; \rho^{(1)}_\phi(x) g(r_1,x) \omega'_\phi(x,r_2) \;
	.
	\label{eq:fields}
\end{align} 
Thus, external fields do not add any complexity to the connectivity problem because the diagrammatical structure remains chain-like and can thus be expressed as a Volterra equation. This unfortunately does not apply for the subject of the next section.

\subsection{Long-ranged Interactions}

\begin{figure}[h!]
	\includegraphics[width=0.49 \textwidth]{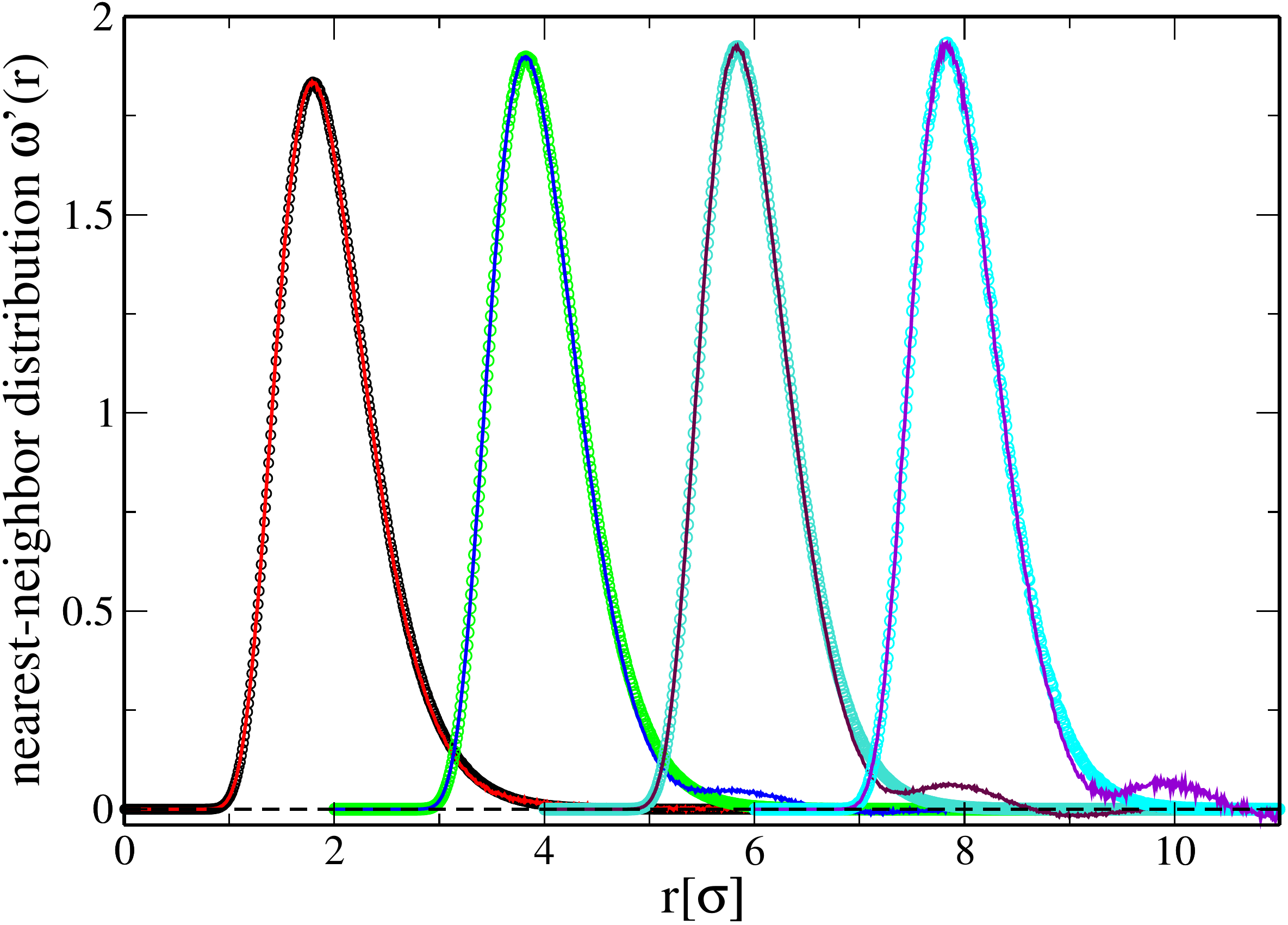} 
	\caption{Nearest-neighbor distributions of systems interacting through the inverse square potential $V_{10}$ including 1,2,5 and 20 neighbors respectively (from left to right, successively shifted by 2d for clarity). Symbols correspond to simulation results whereas lines denote the solution of (\ref{eq:NND}).
		\label{fig:LR_NND}}
\end{figure}

\begin{figure}[h!]
	\includegraphics[width=0.49 \textwidth]{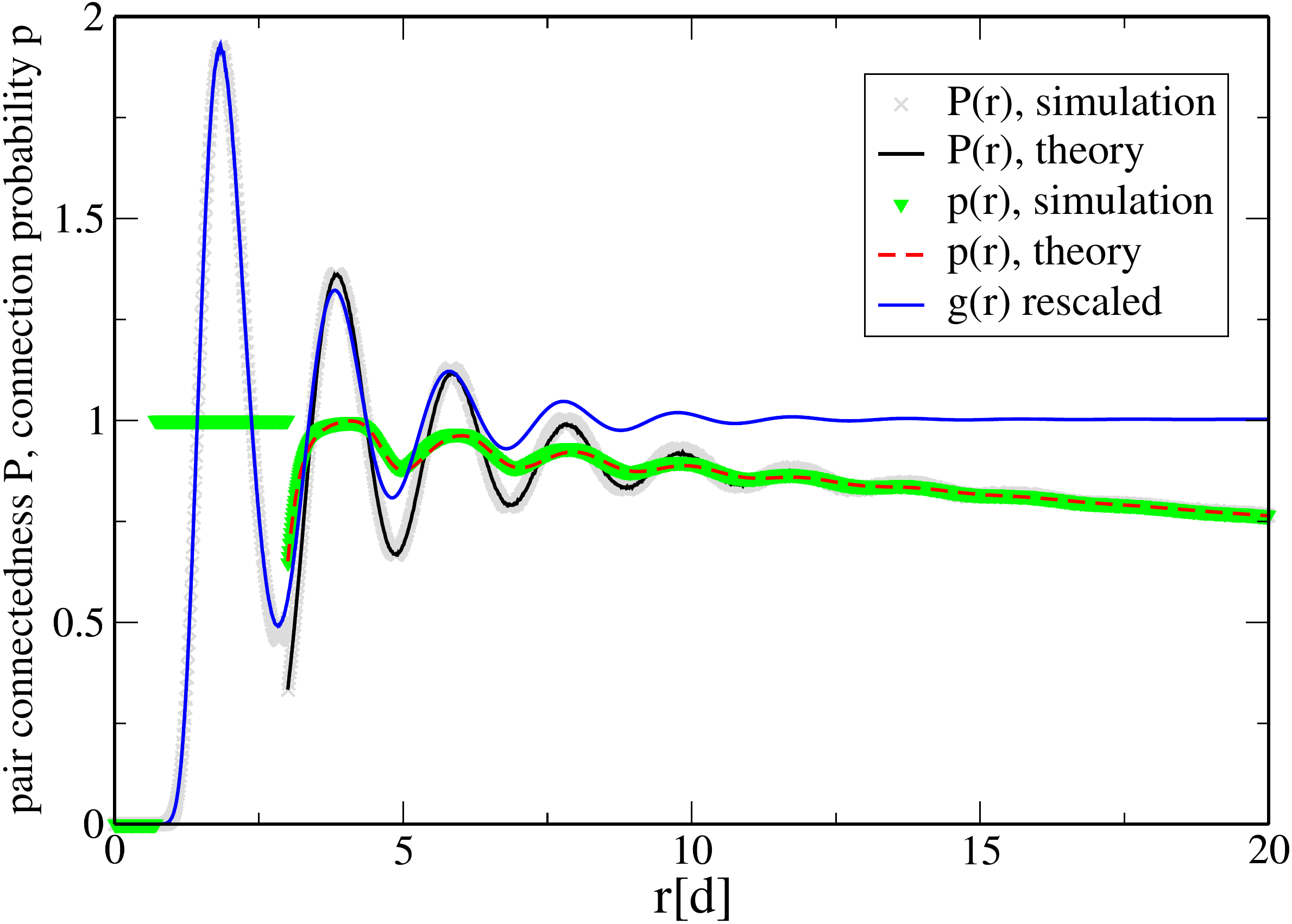} 
	\caption{Pair connectedness and connection probability $p$ for the inverse square potential $V_{10}$ with $\rho = 0.5$, taking into account 5 neighbors. 
		\label{fig:rescaled}}
\end{figure}

In contrast to external fields, for long-range interactions, we cannot ignore three-particle correlations anymore. Thus, there is no straightforward way to determine the nearest-neighbor distribution from just the radial distribution function. But we can at least attempt to characterize the discrepancy. In one-dimension we can order the particles $r_1 < r_2 < ... < r_n$, so that we know  beforehand which particle is neighboring another particle. The system geometry now demands that if particles at $r_1$ and $r_3$ are connected, the same has to apply for $r_1$ and $r_2$ as well as $r_2$ and $r_3$. 
Therefore all diagrams contributing to $P$ share the same carcass, \textit{i.e.} 
\newline
  \begin{center}
  	\begin{tikzpicture}
  	\draw [thick] (1.5,-0.6) circle [radius=0.2];
  	  	  	\node at (1.5,-1.05) {0};
  	\draw [line width=4] (1.7,-0.6) -- (2.3,-0.6);
  	\node at (2.0,-0.9) {$f^\dagger$};
  	\draw [fill] (2.5,-0.6) circle [radius=0.2];
  	  	  	
  	\draw [line width=4] (2.7,-0.6) -- (3.3,-0.6);
  	  	\node at (3.0,-0.9) {$f^\dagger$};
  	\draw [fill] (3.5,-0.6) circle [radius=0.2];
  	  	  	 
  	\draw [line width=4] (3.7,-0.6) -- (4.3,-0.6);
  	  	\node at (4.0,-0.9) {$f^\dagger$};
  	\draw [thick] (5.7,-0.6) circle [radius=0.2];
  	  	  	  	\node at (5.7,-1.05) {$r$};
  	\node at (5,-0.6) {\bf{-}}; 
  	\node at (4.6,-0.8) {\bf{...}}; 
  	\draw [line width=4] (4.9,-0.6) -- (5.5,-0.6);
  	  	\node at (5.2,-0.9) {$f^\dagger$};
  	  		\node at (6.5,-0.9) {.};
  	\end{tikzpicture}
  \end{center}
This diagram is naturally part of eqn.~(\ref{eq:Pexp}), but every addition of an $f^\ast$-bond will result in another diagram of that expansion. Most importantly, all diagrams contributing to $P$ can be constructed in that way. Recall, that our scheme for nearest-neighbor interactions generates chains of the type depicted in fig.~\ref{fig:gw}. 
\newline
 \begin{center}
 	\begin{tikzpicture}
 	\draw [thick] (1.5,-0.6) circle [radius=0.2];
 	\node at (1.5,-1.05) {0};
 	\draw [line width=4] (1.7,-0.6) -- (2.3,-0.6);
 	\node at (2.0,-0.9) {$g^\dagger$};
 	\draw [thick] (2.5,-0.6) circle [radius=0.2];
 	\draw [fill] (2.5,-0.6) circle [radius=0.1];
 	\node at (2.5,-1.05) {$r_1$};
 	\draw [line width=4] (2.7,-0.6) -- (3.3,-0.6);
 	\node at (3.0,-0.9) {$\omega'^\dagger$};
 	\draw [fill] (3.5,-0.6) circle [radius=0.1];
 	\draw [thick] (3.5,-0.6) circle [radius=0.2];
 	\node at (3.5,-1.05) {$r_2$};
 	\draw [line width=4] (3.7,-0.6) -- (4.3,-0.6);
 	\node at (4.0,-0.9) {$\omega'^\dagger$};
 	\draw [thick] (5.7,-0.6) circle [radius=0.2];
 	\node at (5.7,-1.05) {$r$};
 	\node at (5,-0.6) {\bf{-}}; 
 	\node at (4.6,-0.8) {\bf{...}}; 
 	\draw [line width=4] (4.9,-0.6) -- (5.5,-0.6);
 	\node at (5.2,-0.9) {$\omega'^\dagger$};
 	\node at (6.5,-0.9) {.};
 	\end{tikzpicture}
 	\label{fig:cc2}
 \end{center}
At this point we hit two obstacles: On the one hand, for long-ranged interactions $\omega'$, as obtained from eqn.~(\ref{eq:NND}), is not exact, as the pair-distribution hierarchy does not factorize anymore. However, for highly repulsive potentials, the configurations that feature more than one particle within the connectivity shell (to the right) are strongly suppressed energetically. Thus, the additional long-ranged interaction energy hardly alters the short-scale alignment. Figure \ref{fig:LR_NND} illustrates this observation for the inverse square potential and varying interaction ranges. 
\newline
Moreover, it shows that treating the long-ranged interaction pair distribution as a nearest-neighbor interacting system might render $\omega'$ negative - hence $\omega'$ cannot be safely interpreted as a nearest-neighbor distribution for long range interactions. However, since we need input for the thermal distribution anyway, we might as well use the real, \textit{i.e.}~numerically sampled, nearest-neighbor distribution as input. \newline
On the other hand, the hierarchy of neighbor-distribution functions cannot be generated by iteratively convolving $\omega'$. Once the nearest-neighbor is found, we cannot shift the system to its position and expect the same distribution of nearest neighbors to apply for that particle due to the correlation to the previous origin. Essentially, the $P(\tau,r)$ on the right hand side of eqn.~(\ref{eq:Main}) is not the regular pair connectedness anymore but rather the pair connectedness under the constraint that there is a particle already placed at the origin. At this point it is useful to formulate eqn.~(\ref{eq:Main}) in terms of probabilities to be able to invoke Bayes' theorem
\begin{align}
p(0,r) =& \Theta(d-r) \;  \nonumber \\  +& \Theta(r-d) \rho \int_0^d \dd \tau \; \omega'(0,\tau) p(\tau,r | 0) \frac{g(\tau,r)}{g(0,r)} \; .
\label{eq:MainProb}
\end{align}
Since $p$ is a proper probability, we can treat the constrained probability in the integral according to Bayes theorem
\begin{align}
	p(\tau,r | 0) = \frac{p(0|\tau,r )}{p(0)} p(\tau,r) \; .
	\label{eq:Bayes2}
\end{align}
Thus, we can formally write down a Volterra equation for the probability of particles at 0 and $r$ belonging to the same cluster
\begin{align}
	p(0,r) =& \rho \int_{r-d}^{d} \dd x \; \omega'(x,r)\frac{g(0,x)}{g(0,r)} \; +  \;  \nonumber \\   &\rho \int_d^r \dd \tau \; \frac{\omega'(0,\tau) g(\tau,r)}{g(0,r)}  \frac{p(0|\tau,r )}{p(0)} p(\tau,r) \; .
	\label{eq:Pres}
\end{align}
Accordingly, the conditional probability can be absorbed into the kernel. This is hardly surprising as we expect an equation structurally similar to eqn.~(\ref{eq:COZ}). Notice also that we need the complete pair-distribution function as input where previously only the $g(r)$ within the first connectivity was required.
 It turns out, that if we assume statistical independence of events in eqn.~(\ref{eq:Bayes2}) we can already reproduce the pair connectedness for long-range interacting systems to surprisingly high precision (see fig.~\ref{fig:rescaled}).
Equation~(\ref{eq:Pres}) serves as an excellent approximation even with the crudest assumption for the constrained connectivity probability. In this approximation, we essentially compute $P$ as before for nearest-neighbor interactions but also normalize it by the $g$ that is caused by $\omega'$ if we assume only nearest-neighbor interactions, \textit{i.e.} eqn. (\ref{eq:NNDinv}), to get the probability $p$. The substitute $g$ is also shown in figure \ref{fig:rescaled}. By multiplying the probability with the sampled $g$, we get a \textit{rescaled} pair connectedness, which is in excellent agreement with the simulations results.

\subsection{Higher Dimensions}

Finally we sketch the application of the proposed scheme to a three dimensional system. The fundamental difference to the one-dimensional case is that there is more than one path that can lead to a connected configuration. To fight at one front at a time, we will only consider the three-dimensional ideal gas, so that we do not need to worry about correlations of particle positions. As the factorization of the hierarchy of density distributions (eqn.~(\ref{eq:Kirkwood})) holds, the nearest-neighbor distribution can still be determined with an analogue of eqn.~(\ref{eq:NND}). The notion of a nearest-neighbor remains valid and the corresponding distribution function depends exclusively on the distance to a chosen particle. However, since the system does not allow for global order, the idea of acquiring the next-nearest-neighbor distribution through a three-dimensional convolution of nearest-neighbor distributions does not work anymore. We can however switch to eqn.~(\ref{eq:NNDinv}) as the nearest-neighbor distribution in that expression appears only with respect to the origin. As a consequence, we find
\begin{align}
\omega'({\boldsymbol 0},{\boldsymbol r}) &=& &g({\boldsymbol 0},{\boldsymbol r}) -  \rho \int \dd^3 {\boldsymbol x} \; \Theta(r-|{\boldsymbol x}|) \omega'({\boldsymbol 0},{\boldsymbol x}) g({\boldsymbol x},{\boldsymbol r}) \; \nonumber \\
&=& &1 - \rho \int \dd^3 {\boldsymbol x} \;  \Theta(r-|{\boldsymbol x}|) \omega'(|{\boldsymbol x}|)
\label{eq:NND3d}
\end{align}
which is readily solved to yield 
\begin{align}
\omega'({\boldsymbol 0},{\boldsymbol r })= \exp\left(-\frac{4}{3} \pi \rho |{\boldsymbol r}|^3\right) \; .
\end{align}
This expression is still defined on $\mathbb{R}^3$ and normalized accordingly, thus the result coincides with an the expression derived, for instance, by Torquato \cite{torquato1990nearest} once the angular dependencies are integrated out.

The three-dimensional analogue to eqn.~(\ref{eq:MainProb}) for the three-dimensional ideal gas reads
\begin{align}
p(\bs{0},\bs{r}) =& \Theta(d-|{\boldsymbol r}|)  \;  \nonumber \\  +& \Theta(|{\boldsymbol r}|-d) \rho \int \dd^3 {\boldsymbol \tau} \; \Theta(d-|{\boldsymbol \tau}|)  \; \omega'({\boldsymbol 0},{\boldsymbol \tau}) p({\boldsymbol \tau},{\boldsymbol r} | \bs{0}) \; .
\label{eq:Main3D}
\end{align}
In order to make further progress, we need to find a way to treat the conditional probability. First we make use of Bayes theorem again to convert the equation into a standard type integral equation.
\begin{align}
p(\bs{0},\bs{r}) &= \Theta(d-|{\boldsymbol r}|) +\Theta(|{\boldsymbol r}|-d) \times \;  \nonumber \\  &\times \rho \int \dd^3 {\boldsymbol \tau} \; \Theta(d-|{\boldsymbol \tau}|)  \; \omega'({\boldsymbol 0},{\boldsymbol \tau}) \frac{p(\bs{0}|\bs{\tau},\bs{r})}{p( \bs{0})} p({\boldsymbol \tau},{\boldsymbol r}) \; .
\label{eq:3DBayes}
\end{align}
Capitalizing on the homogeneity of the system, we choose  ${\boldsymbol r}=(0,0,r)^t$ and parameterize the position of 
the nearest neighbor  by ${\boldsymbol \tau}=\tau(\cos(\phi)\sin(\vartheta),\sin(\phi)\sin(\vartheta),\cos(\vartheta))^t$. On top of that, the unconditional pair connectedness depends exclusively on the distance between the two arguments so that we obtain
\begin{align}
&p(r) = \Theta(d-r)  +  2 \pi \rho \, \Theta(r-d) \nonumber   \int_0^d \dd \tau \int_0^\pi \dd \theta   \; \\    &  \tau^2 \sin(\theta) \exp\left(-\frac{4}{3} \pi \rho \tau^3 \right) \frac{p(\bs{0}|\bs{\tau},\bs{r})}{p(\bs{0})} p(|\bs{r}-\bs{\tau}|)    \; .
\end{align}
Introducing $\eta:= \frac{\pi}{6}\rho d^3$, the volume fraction of connectivity shells, and substituting $t:=\frac{\tau}{d}$ results in
\begin{align}
&p(r) = \Theta(d-r)  \;  \nonumber + 12 \eta \; \Theta(r-d)  \int_0^1 \dd t\;  t^2 \exp\left(-8 \eta t^3 \right) \\ &\int_0^\pi \dd \theta \;  \sin(\theta)  \frac{p(\bs{0}|\bs{\tau},\bs{r})}{p(\bs{0})} p\left(\sqrt{r^2+ d^2 t^2 -2 r d t \cos(\theta) }\right) \; .
\end{align}
Finally, we replace the angular integral by an integration over $u:=|\bs{\tau}-\bs{r}|$ which importantly depends monotonically on $\theta$ retrieving the familiar form
\begin{align}
	&p(r) = \Theta(d-r)  \;  \nonumber + \frac{12 \eta}{d r} \; \Theta(r-d)  \int_0^1 \dd t\;  t \exp\left(-8 \eta t^3 \right) \\ &\int_{r-d t}^{r+d t} \dd u \;  u \; \frac{p(\bs{0}|\bs{\tau},\bs{r})}{p(\bs{0})}  p(u) \; .
	\label{eq:IDG_main}
\end{align} 
This is the moment we have to leave the realms of exactness to make further progress as we do not know the analytic structure of the correlation
\begin{align}
c(r,t,u) := \frac{p(\bs{0}|\bs{\tau},\bs{r})}{p(\bs{0})} \;.
\label{eq:smallc}
\end{align}

In order to find sensible approximations we need to understand the function appearing in the enumerator of (\ref{eq:smallc}). 
It is the probability that there is a particle at the origin with a spherical cavity of radius $\tau$ around it devoid of particles given particles at $\bs{\tau}$ and $\bs{r}$ which belong to the same cluster. Normalized by the probability of just finding a particle at the origin with no other particles in the open ball $B_\tau(\bs{0})$ we end up with $c(r,t,u)$. However, thanks to the ideal gas, the particle positions are entirely uncorrelated so that the only meaningful information in the condition is the fact that $\bs{\tau}$ and $\bs{r}$ belong to the same cluster. Now, if $|\bs{\tau}-\bs{r}| < d$, even this piece of information is obsolete, meaning 
\begin{align}
c(r,t,u < d) = \frac{p(\bs{0}|\bs{\tau},\bs{r})}{p(\bs{0})} = \frac{p(\bs{\tau},\bs{r}|\bs{0})}{p(\bs{\tau},\bs{r})} = 1 \; .
\label{eq:IDG_Cond1}
\end{align}
The only configurations making a difference are those where $\bs{\tau}$ and $\bs{r}$ are necessarily connected through the  particle at the origin. In other words, $c$ differs from 1 due to all configurations for which the nearest neighbor turns out to be a deadlock on the path to $\bs{r}$. However, $c$ is not always larger than 1 due to the obstructing void that comes with the nearest-neighbor condition. On the contrary, as $t$ approaches unity we expect $c$ to be smaller than 1 because in the limit of $t \rightarrow 1$, the probability of the origin having another neighbor vanishes. As a consequence, the origin only blocks the intersect with the particle at $\bs{\tau}$ for other other potential connectors. Thus,
\begin{align}
c(r,t\rightarrow 1,u )  \leq 1 \; .
\label{eq:IDG_Cond2}
\end{align}
   
On the other hand, if $t$ approaches zero, \textit{i.e.} the  nearest neighbor becomes the origin, the origin again is obsolete and hence
\begin{align}
	c(r,t\rightarrow 0,u) \rightarrow 1 \; .
	\label{eq:IDGCond2}
\end{align}
At this point, we need to close the integral equation by specifying a functional form of $c$. In contrast to the closures typically used in liquid state theory, equation (\ref{eq:IDG_main}) allows for a purely geometrical treatment. 
\begin{figure}
	\begin{center}
		\includegraphics[width=0.45 \textwidth]{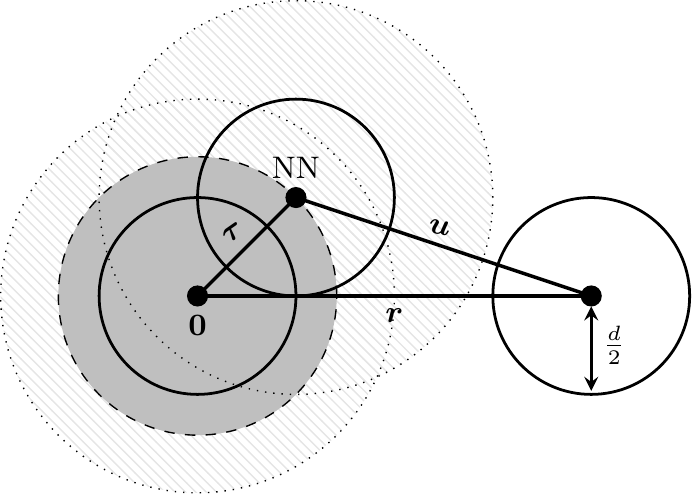}
	\end{center}
	\caption{Illustration of approximation (\ref{eq:Vol}). The hatched region indicates the volume in which additional particles would extend the cluster the origin is part of. The gray area is excluded as the location of the nearest-neighbor (NN) imposes a region devoid of particles around the origin.}
\end{figure}
We assume, the probability of a structure to be part of the same cluster as a distant particle is proportional to the volume it provides for other particles to attach, \textit{i.e.}
\begin{align}
	\tilde{c}(r,t,u) &=& &\frac{p(\bs{\tau},\bs{r}|\bs{0})}{p(\bs{\tau},\bs{r})} \propto \frac{Vol \left[ 	\left( B_{d}(\bs{0}) \cup B_{d} (\bs{\tau}) \right) \setminus B_{\tau}(\bs{0}) \right] }{Vol\left[ B_{d} (\bs{\tau}) \right]} \nonumber \\ 
	&=&  &1 + \frac{3}{4} t -\frac{17}{16} t^3
	\label{eq:Vol}
\end{align}
for $u \geq d$. In order for eq. (\ref{eq:IDGCond2}) to apply, the proportionality constant even has to be unity. Thus, the full function reads:
\begin{align}
	c(t,u) =  \theta(d-u) + \theta(u-d) \left[ 1 + \frac{3}{4} t -\frac{17}{16} t^3 \right]
	\label{eq:capprox}
\end{align}
Apparently, condition (\ref{eq:IDG_Cond1}) is satisfied by this choice for $c$ as well. Naturally, the above expression is an approximation as the actual $c$ has to depend on $r$ and $u$ in a non-trivial way. Nevertheless, plugging (\ref{eq:capprox}) into equation (\ref{eq:IDG_main}) already yields good agreement with simulation results. 
\begin{figure}
	\includegraphics[width=0.49 \textwidth]{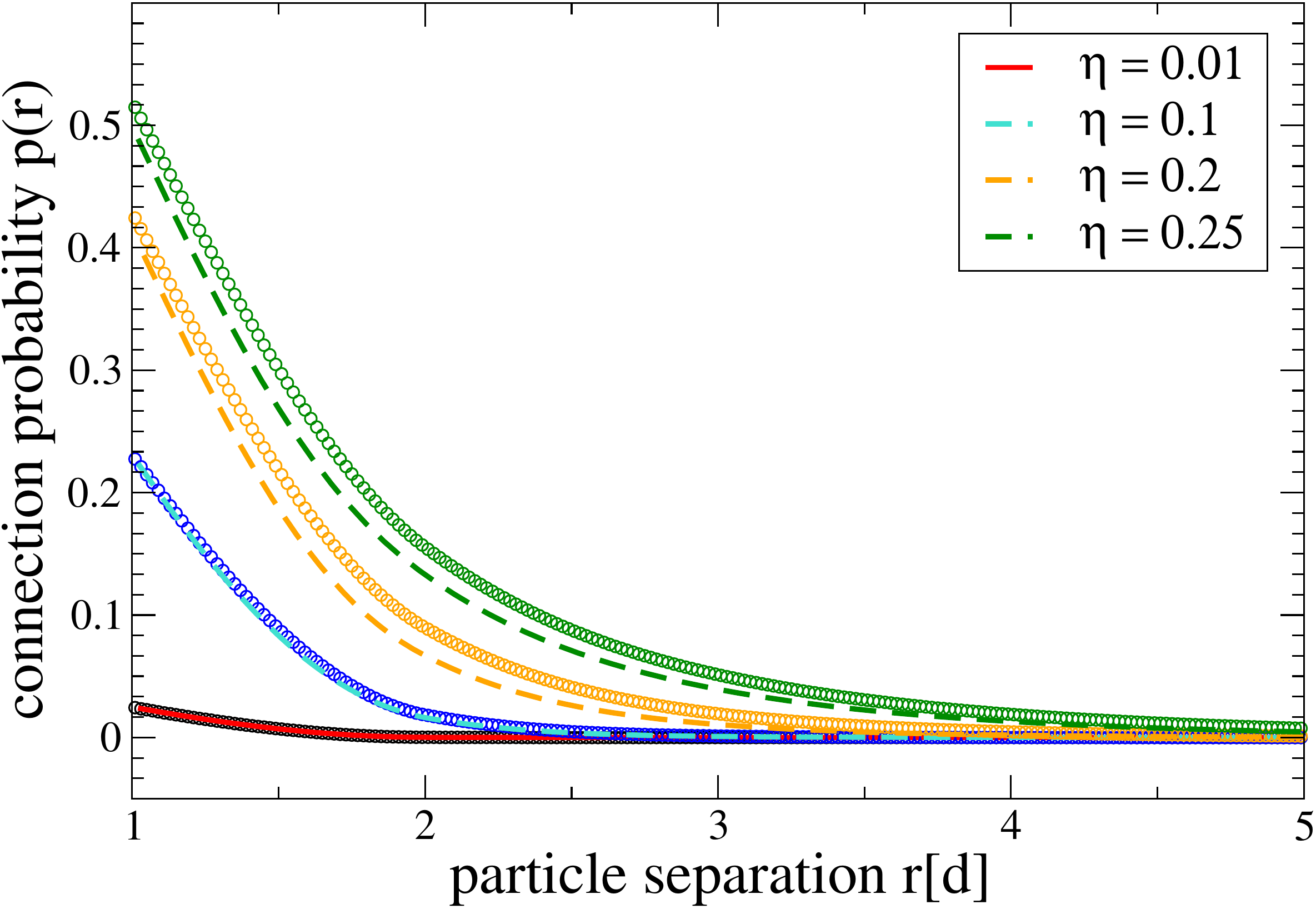}
	\caption{Connection probability $p$ for the three-dimensional ideal gas for different volume fractions. Dashed lines correspond to the numerically determined solution of eq. (\ref{eq:IDG_main}) using the closure (\ref{eq:capprox}) - symbols indicate the associated simulation results.
		\label{fig:idg}}
\end{figure}
\begin{figure}
	\includegraphics[width=0.49 \textwidth]{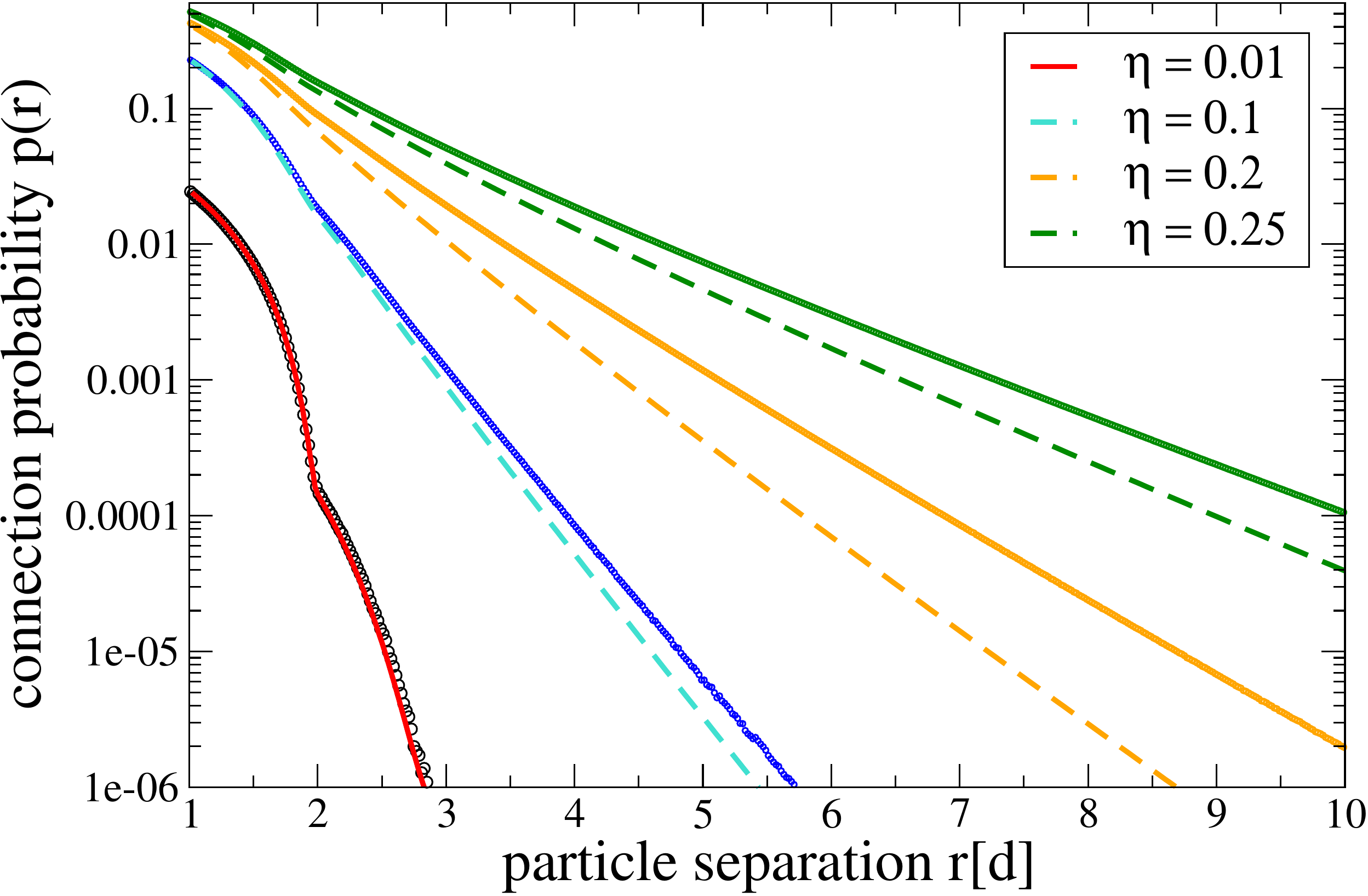}
	\caption{Logarithmic representation of the data shown in fig. \ref{fig:idg}.
		\label{fig:idg2}}
\end{figure}
The above kernel has the primary advantage that due to the simple $u$-dependence, the integrals over $u$ and $t$ can be exchanged in order to yield a $t$-integral which can be performed analytically. Thus,  eq. (\ref{eq:IDG_main}) can be reduced to a simpler form  containing only a single integral which can be considered as a Fredholm integral equation as long as $r$ decays fast enough that $p(r > \xi) = 0$ for a finite $\xi$ is a sensible approximation. As typical for Fredholm equations we can make use of the Picard iteration to solve (\ref{eq:IDG_main}) with the specified kernel, the results are illustrated in figures \ref{fig:idg} and \ref{fig:idg2}. Keeping in mind the crudeness of the employed approximation, the agreement to our simulation results is remarkably good especially for small volume fractions. In the limit of infinite dilution, the probability of a particle having two neighbors is already heavily suppressed, so that the nearest-neighbor being a deadlock in a connected configuration is effectively impossible. As a consequence $c(r,t,u) = 1$ becomes exact for $\eta \rightarrow 0$. For finite volume fractions, the closure asymptotically generates exponential decays which systematically slightly undershoot the simulation results. However, for volume fractions $\eta \gtrsim 0.3$ numerical stability breaks down which is to be expected from Fredholm equations if the integral norm of the kernel becomes too large \cite{tricomi1985integral}. This behavior is independent of the percolation transition as even exactly at the percolation threshold, the designated power-law solution would still belong to the class $L^2$. Additionally, equation (\ref{eq:IDG_main}) can easily be closed to yield solutions decaying like fractals which are perfectly stable numerically. The instability is hence simply an artifact of the closure. There are multiple ways to improve on our choice of $c$ for instance by implementing the geometrical nuances in a less crude fashion. Moreover, since the Picard iteration is computationally not expensive, even brute force methods come to mind. Ultimately, the constrained probability can also be determined by simulations which is as expensive as measuring $p(r)$ right away, but might still be useful as a starting point for more elaborate approximate schemes. However, as this section was supposed to just illustrate, that the range of validity of our derived integral equation exceeds the one-dimensional, further extensions shall be discussed elsewhere.

\section{Conclusion}
We have presented a general method to solve the connectivity problem for one-dimensional systems with an arbitrary nearest-neighbor interaction exactly, given the correct pair-distribution function. For these systems, we showed that the derived integral equation is equivalent to the connectivity Ornstein-Zernike equation, however, it substantially simplifies the derivation of the known exact solutions to continuum percolation models. Moreover, the connectivity properties can be inferred completely from thermal distribution functions. This relation would be of immense practical value if generalized to higher dimensions.
For higher-dimensional systems and long-ranged interactions the analogous integral equation still holds, however, it features a constrained probability. In this case, similar to the connectivity Ornstein-Zernike equation, a closure relation is required. Hence one might argue that compared to the standard approach little is gained as approximations are required after all only for a different function. Yet, in contrast to the direct connectivity, the conditional probability appearing in the kernel is an observable of the system. The integral kernel can simply be sampled by Monte Carlo simulation, and especially in view of universality it might reveal useful insight for future approaches. And, as shown for the ideal gas, even approximations based on simple geometrical considerations already lead to decent results.

\label{sec:conclusion}

\begin{acknowledgments}
We acknowledge funding by the German Research Foundation in the project SCHI 853/4-1.
\end{acknowledgments}

\end{document}